\documentclass[11pt]{article}
\usepackage{cite}
\usepackage{mathrsfs}
\usepackage{amssymb,amsfonts,amsmath}
\usepackage{enumerate}
\usepackage{indentfirst}
\usepackage{latexsym,bm}
\usepackage[noend]{algpseudocode}
\usepackage{algorithmicx,algorithm}
\usepackage{algorithm}
\usepackage{makeidx}
\usepackage{fancybox}
\usepackage{color}
\usepackage{multicol}
\usepackage[latin1]{inputenc}
\usepackage{graphicx}
\usepackage{pstricks,pst-node,pst-text,pst-3d}
\usepackage{tikz}
\newtheorem{thm}{Theorem}[section]

\newtheorem{lem}[thm]{Lemma}

\newtheorem{pro}[thm]{Proposition}

\newtheorem{rem}{Remark}[section]
\newenvironment{pf}{{\noindent \it \bf Proof:}}{{\hfill$\Box$}\\}

\def\qed{\hfill \nopagebreak\rule{5pt}{8pt}}

\usetikzlibrary{shapes,snakes}
\newcommand{\AY}[1]{{#1}}
\newcommand{\jbj}[1]{{#1}}

 \newcommand{\GG}[1]{{#1}}

\newcommand{\2}{\vspace{0.2cm}}
\newcommand{\induce}[2]{#1 \langle  #2 \rangle }
\DeclareMathOperator{\ind}{ind}
\newcommand{\claimA}{A}
\newcommand{\claimB}{B}
\newcommand{\claimC}{D}
\newcommand{\claimD}{C}

\newcommand{\dom}{\rightarrow}

\newcommand{\good}{{strong arc}}
\newcommand{\Good}{{Strong arc}}

\newcommand{\redC}[1]{red}
\newcommand{\redT}[1]{0.05cm}
\newcommand{\blueC}[1]{blue}
\newcommand{\blueT}[1]{0.02cm}
\newcommand{\ColorXX}[2]{#1}

\begin{document}

\title{\bf Arc-disjoint Strong Spanning Subdigraphs of Semicomplete Compositions\thanks{Research supported by the Danish research council under grant number DFF-7014-00037B.}}
\author{J\o{}rgen Bang-Jensen$^{1}${ } Gregory Gutin$^{2}$\
{ } Anders Yeo$^{1}${ }\\
$^{1}$ IMADA,
University of Southern Denmark\\
Odense, Denmark\\
jbj@imada.sdu.dk, andersyeo@gmail.com\\
$^{2}$ Department of Computer Science\\
Royal Holloway, University of London\\
Egham, Surrey, TW20 0EX, UK\\
g.gutin@rhul.ac.uk}
\date{}
\maketitle

\begin{abstract}
  A {\bf \good{} decomposition} of a digraph $D=(V,A)$ is a decomposition 
of  its arc set $A$ into two disjoint subsets $A_1$ and $A_2$ such that both of the spanning subdigraphs
$D_1=(V,A_1)$ and $D_2=(V,A_2)$ are strong.
Let $T$ be a digraph with $t$ vertices $u_1,\dots , u_t$ and let $H_1,\dots H_t$ be digraphs such that $H_i$ has vertices $u_{i,j_i},\ 1\le j_i\le n_i.$
Then the composition $Q=T[H_1,\dots , H_t]$ is a digraph with vertex set $\cup_{i=1}^t V(H_i)=\{u_{i,j_i}\mid 1\le i\le t, 1\le j_i\le n_i\}$ and arc set
\[
\left(\cup^t_{i=1}A(H_i) \right) \cup  \left( \cup_{u_iu_p\in A(T)} \{u_{ij_i}u_{pq_p} \mid 1\le j_i\le n_i, 1\le q_p\le n_p\} \right).
\]
We obtain a characterization
of digraph  compositions $Q=T[H_1,\dots H_t]$ which have  a \good{} decomposition when $T$ is a semicomplete digraph and 
each $H_i$ is an arbitrary digraph. Our characterization generalizes a characterization 
by Bang-Jensen and Yeo (2003) of semicomplete digraphs with a \good{} decomposition and 
solves an open problem by Sun, Gutin and Ai (2018) on \good{} decompositions of digraph compositions $Q=T[H_1,\dots , H_t]$
in which $T$ is semicomplete and each $H_i$ is arbitrary. Our proofs are constructive and imply the existence of a polynomial algorithm for constructing a \good{} decomposition of a digraph $Q=T[H_1,\dots , H_t]$, with $T$ semicomplete, whenever such a decomposition exists.
\vspace{0.3cm}\\
{\bf Keywords:} strong spanning subdigraph; decomposition into
strong spanning subdigraphs; semicomplete digraph; digraph composition.
\vspace{0.3cm}\\ {\bf AMS subject
classification (2010)}: 05C20, 05C70, 05C76, 05C85.

\end{abstract}

\section{Introduction}
We refer the reader to \cite{bang2009,bang2018}
 for graph theoretical notation and terminology not given
here. A {\bf digraph} is not allowed to have parallel arcs or loops. 
A {\bf directed multigraph} $D=(V,A)$ can have parallel arcs, i.e. $A$ is a multiset. 
A directed multigraph $D = (V, A)$ is {\bf strongly connected} (or {\bf
strong}) if there exists a path from $x$ to $y$ and a path from $y$
to $x$ in $D$ for every pair of distinct vertices $x, y$ of $D$. A
directed multigraph $D$ is {\bf$k$-arc-strong} if $D- X$ is strong for every
subset $X\subseteq A$ of size at most $k- 1$. 

A directed multigraph $D=(V,A)$ has a {\bf \good{} decomposition} if $A$ \jbj{can be partitioned into} disjoint subsets $A_1$ and $A_2$ such that both
$(V,A_1)$ and $(V,A_2)$ are strong \cite{bangJCTB102}. 
A directed multigraph $D$ is {\bf semicomplete} if there is an arc between any pair of distinct vertices in $D$. 
In particular, a tournament is semicomplete digraph with just one arc between any pair of distinct vertices. 
(A semicomplete digraph can have two arcs between a pair $x,y$ of distinct vertices: $xy$ and $yx$.)

Bang-Jensen and Yeo \cite{bangC24} proved that 
it is NP-complete to decide whether a digraph has a \good{} decomposition. They also characterized semicomplete digraphs with a
\good{} decomposition. Note that every digraph with a \good{} decomposition must be  2-arc-strong.

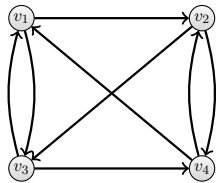
\begin{figure}[!h]
\begin{center}
\tikzstyle{vertexX}=[circle,draw, top color=gray!5, bottom color=gray!30, minimum size=16pt, scale=0.6, inner sep=0.5pt]
\tikzstyle{vertexY}=[circle,draw, top color=gray!5, bottom color=gray!30, minimum size=20pt, scale=0.7, inner sep=1.5pt]

\begin{tikzpicture}[scale=0.4]
 \node (v1) at (1.0,6.0) [vertexX] {$v_1$};
 \node (v2) at (7.0,6.0) [vertexX] {$v_2$};
 \node (v3) at (1.0,1.0) [vertexX] {$v_3$};
 \node (v4) at (7.0,1.0) [vertexX] {$v_4$};
\draw [->, line width=0.03cm] (v1) -- (v2);
\draw [->, line width=0.03cm] (v2) -- (v3);
\draw [->, line width=0.03cm] (v3) -- (v4);
\draw [->, line width=0.03cm] (v4) -- (v1);
\draw [->, line width=0.03cm] (v1) to [out=285, in=75] (v3);
\draw [->, line width=0.03cm] (v3) to [out=105, in=255] (v1);
\draw [->, line width=0.03cm] (v2) to [out=285, in=75] (v4);
\draw [->, line width=0.03cm] (v4) to [out=105, in=255] (v2);
\end{tikzpicture}
\caption{Digraph $S_4$}\label{S4fig}
\end{center}
\end{figure}

\begin{thm}\label{thm03}\cite{bangC24}
A 2-arc-strong semicomplete digraph $D$ has a \good{} decomposition 
if and only if $D$ is not isomorphic to
$S_4$, where $S_4$ is obtained from the complete digraph with four
vertices by deleting the arcs of a cycle of length four (see Figure \ref{S4fig}). Furthermore, a \good{}
decomposition of $D$ can be obtained in polynomial time when it
exists.
\end{thm}

The following result by Bang-Jensen and Huang extends Theorem \ref{thm03} to locally semicomplete digraphs. 
A digraph is {\bf locally semicomplete} if every two vertices with a common out- or in-neighbour have 
an arc between them. Clearly, the class of locally semicomplete digraphs is a generalization of semicomplete digraphs. 

\begin{thm}\label{thm04}\cite{bangJCTB102}
A 2-arc-strong locally semicomplete digraph $D$ has a \good{} decomposition if and only if $D$ is not
the square of an even cycle\footnote{The {\bf square} of a directed cycle $v_1v_2\ldots{}v_nv_1$ is obtained by adding an arc from $v_i$ to $v_{i+2}$ for every $i\in [n]$, where $v_{n+1}=v_1$ and $v_{n+2}=v_2$.}.
\end{thm}

Let $T$ be a digraph with $t$ vertices $u_1,\dots , u_t$ and let $H_1,\dots H_t$ be digraphs such that $H_i$ has vertex set  $\{u_{i,j_i}|1\le j_i\le n_i\}$
Then the {\bf composition} $Q=T[H_1,\dots , H_t]$ is a digraph with vertex set $\cup_{i=1}^t V(H_i)$ and arc set

\[
\left(\cup^t_{i=1}A(H_i) \right) \cup  \left( \cup_{u_iu_p\in A(T)} \{u_{ij_i}u_{pq_p} \mid 1\le j_i\le n_i, 1\le q_p\le n_p\} \right) .
\]


We say that  a composition $Q=T[H_1,\dots , H_t]$ is a
{\bf semicomplete composition} if $T$ is semicomplete. In the important special case when each $H_i$ has no arc we say that $Q$ is an {\bf extension} of $T$. In particular the class of {\bf extended semicomplete digraphs} consists of all digraphs that are extensions of a semicomplete digraph, that is, of the form $Q=T[\overline{K}_{n_1},\ldots{},\overline{K}_{n_t}]$ where $T$ is a semicomplete digraph and  $\overline{K}_r$ is a digraph on $r$ vertices and no arcs.

Recently, Sun, Gutin and Ai \cite{sun} proved the following charaterization of a subset of semicomplete compositions with a \good{} decomposition, where $\overrightarrow{C}_3$
 is a directed cycle on three vertices, $\overline{K}_p$ is a digraph with $p$ vertices and no arcs, and $\overrightarrow{P}_2$ is a directed path on two vertices (that is, it is just an arc).

\begin{thm}\label{good-decompSemi}\cite{sun}
Let $T$ be a strong semicomplete digraph on $t\ge 2$ vertices and let $H_1,\dots ,H_t$ be arbitrary digraphs, each with at least two vertices.
Then $Q=T[H_1,\dots ,H_t]$ has a \good{} decomposition if and only if $Q$ is not isomorphic to one of the following three digraphs: 
$\overrightarrow{C}_3[\overline{K}_2,\overline{K}_2,\overline{K}_2]$, $\overrightarrow{C}_3[\overline{K}_2,\overline{K}_2,\overrightarrow{P}_2],$
$\overrightarrow{C}_3[\overline{K}_2,\overline{K}_2,\overline{K}_3].$  
\end{thm}

\begin{rem}\label{viewrem}
Note that all three exceptions in Theorem \ref{good-decompSemi} are extended semicomplete digraphs (the middle one is an extension of the unique strong tournament \GG{$T^s_4$} on four vertices, see Figure 2.
\end{rem}

\begin{figure}[!htb] \label{fig_gd}
\begin{center}
\tikzstyle{vertexX}=[circle,draw, top color=gray!5, bottom color=gray!30, minimum size=16pt, scale=0.6, inner sep=0.5pt]
\tikzstyle{vertexY}=[circle,draw, top color=gray!5, bottom color=gray!30, minimum size=20pt, scale=0.7, inner sep=1.5pt]
\begin{tikzpicture}[scale=0.45]
\node (p3) at (2.0,3.0) [vertexX] {$u_{3,1}$};
 \node (p2a) at (6.0,5.0) [vertexX] {$u_{2,1}$};
 \node (p2b) at (6.0,1.0) [vertexX] {$u_{2,2}$};
 \node (p1a) at (10.0,5.0) [vertexX] {$u_{1,1}$};
 \node (p1b) at (10.0,1.0) [vertexX] {$u_{1,2}$};
 \node (p0) at (14.0,3.0) [vertexX] {$u_{4,1}$};

\draw [->, line width=0.03cm] (p0) -- (p1a);
\draw [->, line width=0.03cm] (p0) -- (p1b);
\draw [->, line width=0.03cm] (p1a) -- (p2a);
\draw [->, line width=0.03cm] (p1a) -- (p2b);
\draw [->, line width=0.03cm] (p1b) -- (p2a);
\draw [->, line width=0.03cm] (p1b) -- (p2b);
\draw [->, line width=0.03cm] (p2a) -- (p3);
\draw [->, line width=0.03cm] (p2b) -- (p3);

 \draw [->, line width=0.03cm] (p3) to [out=50, in=130] (p1a);
 \draw [->, line width=0.03cm] (p3) to [out=310, in=230] (p1b);

 \draw [->, line width=0.03cm] (p2a) to [out=50, in=130] (p0);
 \draw [->, line width=0.03cm] (p2b) to [out=310, in=230] (p0);

\draw [->, line width=0.03cm] (p3) to [out=70, in=110] (p0);
\end{tikzpicture}
\caption{$T^s_4[\overline{K}_{2},\overline{K}_{2},\overline{K}_{1},\overline{K}_{1}]$}
\end{center}
\end{figure}
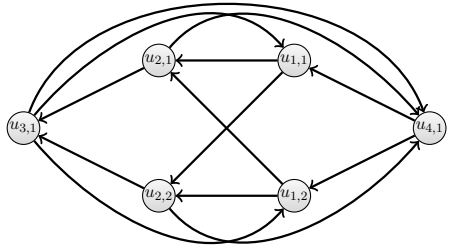
In this paper, solving an open problem in \cite{sun}, we obtain a characterization of {\bf all} semicomplete compositions with a \good{} decomposition.
Note that a digraph with a \good{} decomposition is 2-arc-strong.
Our characterization is as follows:
\begin{thm}\label{mainT}
Let $T$ be a strong semicomplete digraph on $t\ge 2$ vertices and let $H_1,\dots ,H_t$ be arbitrary digraphs.
Then $D=T[H_1,\dots ,H_t]$ has a \good{} decomposition if and only if $D$ is 2-arc-strong and is not isomorphic to one of the following four digraphs: $S_4,$
$\overrightarrow{C}_3[\overline{K}_2,\overline{K}_2,\overline{K}_2]$, $\overrightarrow{C}_3[\overline{K}_2,\overline{K}_2,\overrightarrow{P}_2],$
$\overrightarrow{C}_3[\overline{K}_2,\overline{K}_2,\overline{K}_3].$  
\end{thm}
It is remarkable that all the four exemptions in this theorem are simply the union of the exemptions in Theorems  \ref{thm03} and \ref{good-decompSemi}.
However, we see no simple way to prove our theorem by a direct reduction to Theorems  \ref{thm03} and \ref{good-decompSemi} and do not believe that such a reduction exists. 
Note that the digraphs covered by Theorem \ref{mainT} but not by Theorem \ref{good-decompSemi} are all semicomplete decompositions in which at least one $H_i$ has just one vertex.
Having just one vertex in some $H_i$'s makes the strong arc decomposition problem on semicomplete digraphs \jbj{much}  more complicated \jbj{than} the case when all
$H_i$'s have at least two vertices since in the latter case the semicomplete composition has more symmetries (i.e., authomorphisms) that can be exploited in the proofs. 
Theorem  \ref{thm03} covers just a special subcase of the former case and its proof in \cite{bangC24} is not easier than that of Theorem \ref{good-decompSemi} in \cite{sun}.

Apart from Theorems  \ref{thm03} and \ref{good-decompSemi} used in our proof of Theorem \ref{mainT}, we apply several other results including Edmonds' branching theorem, the existence of nice vertex decompositions proved in \cite{bangEulerpaper} (for details see the next section) and an extension of Theorem \ref{thm03} to directed multigraphs (Theorem \ref{sdmulti}) proved in this paper. Interestingly, the extension of Theorem \ref{thm03} has three further exceptions.

Note that the class of strong semicomplete compositions is a generalization of strong quasi-transitive digraphs by the following recursive characterization of quasi-transitive 
digraphs by Bang-Jensen and Huang \cite{bangJGT20}. A digraph $D=(V,A)$ is {\bf quasi-transitive} ({\bf transitive}) if for any triple $x,y,z$ of distinct vertices of $D$, $xy,yz\in A$ implies that there is an arc between $x$ and $z$ (from $x$ to $z$). Clearly, the class of quasi-transitive digraphs is a generalization of semicomplete digraphs.
For a recent overview of quasi-transitive digraphs and their generalization, see \cite{galeana2018}. 
\begin{thm}\label{qtdchar}\cite{bangJGT20}
Let $D$ be a quasi-transitive digraph. 
\begin{description}
\item[(a)] If $D$ is strong, then there exists  a strong semicomplete
digraph $S$ with $s$ vertices and
quasi-transitive digraphs $Q_1,Q_2,\ldots{},Q_s$ such that $Q_i$
is either a vertex or is  non-strong and
$D=S[Q_1,Q_2,\ldots{},Q_s]$. 
\item[(b)] If $D$ is not strong,
then there exist a transitive oriented graph $T$ with $t$ vertices
and strong quasi-transitive digraphs
$H_1,H_2,\ldots ,H_t$ such that $D=T[H_1,H_2,\ldots ,H_t]$.
\end{description}
\end{thm}

Theorem \ref{mainT} implies a characterization of quasi-transitive digraphs with a \good{} decomposition \GG{(this solves another open question in \cite{sun}).} 
In fact the following follows immediately from Theorems~\ref{qtdchar} and \ref{mainT}
\GG{(observe also that all four exceptions in Theorem \ref{mainT} are quasi-transitive digraphs).
\begin{thm}\label{mainQT}
Let $D$ be a quasi-transitive digraph.  $D$ has a \good{} decomposition if and only if $D$ is 2-arc-strong and 
is not isomorphic to one of the following four digraphs: $S_4,$
$\overrightarrow{C}_3[\overline{K}_2,\overline{K}_2,\overline{K}_2]$, $\overrightarrow{C}_3[\overrightarrow{P}_2,\overline{K}_2,\overline{K}_2],$
$\overrightarrow{C}_3[\overline{K}_2,\overline{K}_2,\overline{K}_3].$  
\end{thm}

To see that strong quasi-transitive digraphs form a  relatively small subset of strong semicomplete compositions, note that the Hamiltonicity problem is polynomial-time solvable for quasi-transitive digraphs \cite{gutinAJC10}, but observed to be NP-complete for strong semicomplete compositions \cite{ai}.

The paper is organized as follows. The next section provides additional terminology and notation and and a number of results used later in the paper. We prove an extension of Theorem \ref{thm03} to directed multigraphs in Section \ref{sec:multi}. In Section \ref{sec:cutvertex} we prove a lemma which simplifies our further proofs: 
every 2-arc-strong semicomplete composition containing a cut-vertex has a strong arc
decomposition. Our main result, Theorem \ref{mainT}, is proved in Section \ref{sec:main}.
However, the proof of Theorem \ref{mainT}  uses our main technical result, Theorem \ref{mainX}, which is proved in Section \ref{sec:X}. We complete the paper in Section \ref{sec:disc}, where we briefly
discuss some open problems.
}

\section{Additional Terminology, Notation and Results}\label{sec:term}

Let $D=(V,A)$ be a directed multigraph.
The {\bf mutiplicity}, $\mu{}(x,y)$ of an arc $xy$ in  $D$ is the number of copies of $xy$ in $D$. An arc is {\bf  single} ({\bf  double}, respectively) if it is of multiplicity 1 (2, respectively).
\GG{For $S\subset V$ such that $S\neq \emptyset$, let 
$(S,T)_D$ be the set of arcs of $D$ with tails in $S$ and heads in $T,$ where $T=V-S.$
The sets of tails in $S$ and heads in $T$ of arcs in $(S,T)_D$ are denoted by $N_D^-(T)$ and $N_D^+(S),$ respectively.
The cardinalities of $N_D^-(T)$ and $N_D^+(S)$ are denoted by $d^-_D(T)$ and $d^+_D(S),$ respectively.}

If $X$ and $Y$ are disjoint vertex sets in a digraph, then we use the notation $X\dom Y$ to denote that $xy$ is an arc for every choice of $x\in X,y\in Y$.
For a non-empty subset $X$ of $V$, the subdigraph of $D$ induced by $X$ is denoted by $\induce{D}{X}.$
Let $P=x_1x_2\dots x_p$ be a path in $D$. For $1\le i\le j\le p$, $P[x_i,x_j]=x_ix_{i+1}\dots x_j$ denotes the subpath of $P$ from $x_i$ to $x_j$.
\jbj{An arc $uv$ of a digraph $D=(V,A)$ is a {\bf cut-arc} if $D-uv$ is not strongly connected. A vertex $v\in V$ is a {\bf cut-vertex} if $D-v$ is not strongly connected.}

A {\bf  vertex decomposition of a digraph }$D$ is a partition $(S_1, \dots , S_p)$, $p\geq 1$, of its vertex set.
The {\bf  index} of vertex $v$ in the decomposition, denoted by $\ind(v)$, is the integer $i$ such that $v\in S_i$.
An arc $uv$ is {\bf  forward} if $\ind(u) < \ind(v)$, {\bf  backward} if $\ind(u) > \ind(v)$.
A vertex decomposition $(S_1,\ldots{},S_p)$ is {\bf  strong} if $\induce{D}{S_i}$ is strong for all $1\leq i\leq p$.
A {\bf  nice vertex decomposition} of a digraph $D$ is a strong decomposition such that the set of cut-arcs of $D$ is exactly the set of backward arcs.

\begin{thm}\label{prop:nice}\cite{bangEulerpaper}
Every strong semicomplete digraph of order at least $4$ admits a nice decomposition.
\end{thm}

\jbj{
\begin{pro}\label{prop:nosame}\cite{bangEulerpaper}
Let $(S_1, \dots, S_p)$ be a nice decomposition of a strong semicomplete digraph $D$.
The following properties hold:
\begin{itemize}
 \item[(i)] If $u_1v_1$ and $u_2v_2$ are two cut-arcs, then $\ind(u_1)\neq \ind(u_2)$ and $\ind(v_1)\neq \ind(v_2)$.
\item[(ii)] If $\ind(u_1)<\ind(u_2)$ then $\ind(v_1)<\ind(v_2)$.
\end{itemize}
\end{pro}}

The following simple lemma sometimes allows one to reduce the number of digraphs under consideration in proofs of results on strong arc decompositions. 

\begin{lem}\label{lemo}\cite{sun}
Let $D=Q[H_1,\dots , H_t],$ where $D$ is an arbitrary digraph and every $H_i$ has no arcs. 
If an induced subdigraph $D'$ of $D$ with at least one vertex in each $H_i$
has a \good{} decomposition, then so has $D.$
\end{lem}

\begin{lem}
  \label{extended2AS}
  Let $D=R[\overline{K}_{n_1},\ldots{},\overline{K}_{n_r}]$ be an extension of a digraph $R$. If $D$ is 2-arc-strong and some $n_i$ is larger than 2, then the digraph $D'$ obtained from $D$ by deleting a vertex from $\overline{K}_{n_i}$ is also 2-arc-strong.
\end{lem}
\begin{pf} Let $x$ be the vertex that we deleted from $H_i=\overline{K}_{n_i}$ and let $y,z$ be two other vertices of $H_i$. 
Suppose that $D'$ is not 2-arc-strong. 
Then there exists a vertex partition $(X,\overline{X})$ of $V(D')$ so that there is at most one arc from $X$ to $\overline{X}$ in $D'$. 
As $D$ is 2-arc-strong this implies that $x$ has an out-neighbour $w^+$ in $\overline{X}$ and an in-neighbour $w^-$ in $X$. 
However now $w^- y w^+$ and $w^- z w^+$ are two arc-disjoint paths from $X$ to $\overline{X}$ in $D'$, contradicting the fact
that there is at most one arc from $X$ to $\overline{X}$ in $D'$.
Hence $D'$ is 2-arc-strong.
\end{pf}

An {\bf  out-branching} ({\bf  in-branching}, resp.) $B$ {\bf rooted} at vertex $z$ in a directed multigraph $D$ is a spanning subdigraph, which is an oriented tree such that only $z$ has in-degree (out-degree, resp.) zero. A vertex of an out-branching (in-branching, resp.) is called a {\bf leaf} if its out-degree (in-degree, resp.) equals zero.
We will use the following result called Edmonds' branching theorem.

\begin{thm}\label{Edmthm}\cite{edmonds1973}
A directed multigraph $D=(V,A)$ with a vertex $z$, has $k$ arc-disjoint out-branchings rooted at $z$ if and only if $d^-(X)\ge k$ for all non-empty $X\subseteq V\setminus \{z\}.$ 
\end{thm}
Note that, by Menger's theorem, the condition of Theorem \ref{Edmthm} is equivalent to the existence of $k$ arc-disjoint paths from $z$ to any vertex $x\in V\setminus \{z\}.$

\section{Extending Theorem \ref{thm03} to Semicomplete Directed Mutigraphs} \label{sec:multi} 

If the arc $xy$ in a directed multigraph $D$ has multiplicity $\mu{}(x,y)\ge 3$, we may delete $\mu-2$ copies of $D$ and the resulting 
directed multigraph has a \good{} decomposition if and only if so has $D$. 
Thus, we may assume that all directed multigraphs considered in this paper have no arcs of multiplicity 3 or more. 

Recall the semicomplete digraph $S_4$ from Theorem \ref{thm03}. Without loss of generality, we may assume that $V(S_4)=\{v_1,v_2,v_3,v_4\}$ and $A(S_4)=\{v_1v_2,v_2v_3,v_3v_4,v_4v_1, v_1v_3,v_3v_1,v_2v_4,v_4v_2\}.$ 
We call the cycle $v_1v_2v_3v_4v_1$ the {\bf  Hamilton cycle} of $S_4$ and cycles $v_1v_3v_1$ and $v_2v_4v_2$ {\bf  2-cycles} of $S_4$.

To prove Theorem \ref{sdmulti}, we will use the following lemma.

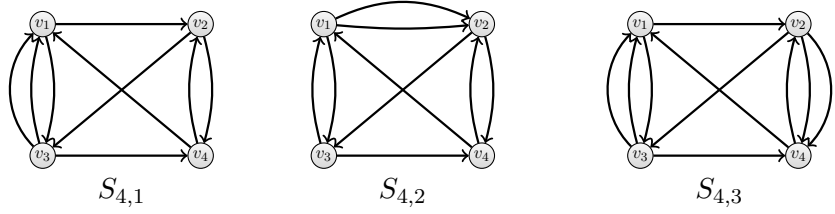
\begin{figure}[!h]
\begin{center}
\tikzstyle{vertexX}=[circle,draw, top color=gray!5, bottom color=gray!30, minimum size=16pt, scale=0.6, inner sep=0.5pt]
\tikzstyle{vertexY}=[circle,draw, top color=gray!5, bottom color=gray!30, minimum size=20pt, scale=0.7, inner sep=1.5pt]
\begin{tikzpicture}[scale=0.35]
 \node (v1) at (1.0,6.0) [vertexX] {$v_1$};
 \node (v2) at (7.0,6.0) [vertexX] {$v_2$};
 \node (v3) at (1.0,1.0) [vertexX] {$v_3$};
 \node (v4) at (7.0,1.0) [vertexX] {$v_4$};
\draw [->, line width=0.03cm] (v1) -- (v2);
\draw [->, line width=0.03cm] (v2) -- (v3);
\draw [->, line width=0.03cm] (v3) -- (v4);
\draw [->, line width=0.03cm] (v4) -- (v1);
\draw [->, line width=0.03cm] (v1) to [out=285, in=75] (v3);
\draw [->, line width=0.03cm] (v3) to [out=105, in=255] (v1);
\draw [->, line width=0.03cm] (v3) to [out=135, in=225] (v1);
\draw [->, line width=0.03cm] (v2) to [out=285, in=75] (v4);
\draw [->, line width=0.03cm] (v4) to [out=105, in=255] (v2);
\node at (4,-0.4) {$S_{4,1}$};
\end{tikzpicture}\hspace{1.1cm}
\begin{tikzpicture}[scale=0.35]
 \node (v1) at (1.0,6.0) [vertexX] {$v_1$};
 \node (v2) at (7.0,6.0) [vertexX] {$v_2$};
 \node (v3) at (1.0,1.0) [vertexX] {$v_3$};
 \node (v4) at (7.0,1.0) [vertexX] {$v_4$};
\draw [->, line width=0.03cm] (v1) to [out=25, in=155] (v2);
\draw [->, line width=0.03cm] (v1) to [out=355, in=185]  (v2);
\draw [->, line width=0.03cm] (v2) -- (v3);
\draw [->, line width=0.03cm] (v3) -- (v4);
\draw [->, line width=0.03cm] (v4) -- (v1);
\draw [->, line width=0.03cm] (v1) to [out=285, in=75] (v3);
\draw [->, line width=0.03cm] (v3) to [out=105, in=255] (v1);
\draw [->, line width=0.03cm] (v2) to [out=285, in=75] (v4);
\draw [->, line width=0.03cm] (v4) to [out=105, in=255] (v2);
\node at (4,-0.4) {$S_{4,2}$};
\end{tikzpicture} \hspace{1.1cm}
\begin{tikzpicture}[scale=0.35]
 \node (v1) at (1.0,6.0) [vertexX] {$v_1$};
 \node (v2) at (7.0,6.0) [vertexX] {$v_2$};
 \node (v3) at (1.0,1.0) [vertexX] {$v_3$};
 \node (v4) at (7.0,1.0) [vertexX] {$v_4$};
\draw [->, line width=0.03cm] (v1) -- (v2);
\draw [->, line width=0.03cm] (v2) -- (v3);
\draw [->, line width=0.03cm] (v3) -- (v4);
\draw [->, line width=0.03cm] (v4) -- (v1);
\draw [->, line width=0.03cm] (v1) to [out=285, in=75] (v3);
\draw [->, line width=0.03cm] (v3) to [out=105, in=255] (v1);
\draw [->, line width=0.03cm] (v3) to [out=135, in=225] (v1);
\draw [->, line width=0.03cm] (v2) to [out=315, in=45] (v4);
\draw [->, line width=0.03cm] (v2) to [out=285, in=75] (v4);
\draw [->, line width=0.03cm] (v4) to [out=105, in=255] (v2);
\node at (4,-0.4) {$S_{4,3}$};
\end{tikzpicture}
\caption{The digraphs $S_{4,1}$, $S_{4,2}$, $S_{4,3}$}\label{S4likefig}
\end{center}
\end{figure}

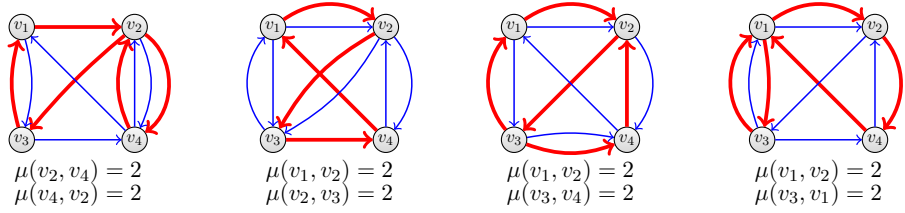
\begin{figure}[!h]
\begin{center}
\tikzstyle{vertexX}=[circle,draw, top color=gray!5, bottom color=gray!30, minimum size=16pt, scale=0.6, inner sep=0.5pt]
\tikzstyle{vertexY}=[circle,draw, top color=gray!5, bottom color=gray!30, minimum size=20pt, scale=0.7, inner sep=1.5pt]
\begin{tikzpicture}[scale=0.3]
 \node (v1) at (1.0,6.0) [vertexX] {$v_1$};
 \node (v2) at (6.0,6.0) [vertexX] {$v_2$};
 \node (v3) at (1.0,1.0) [vertexX] {$v_3$};
 \node (v4) at (6.0,1.0) [vertexX] {$v_4$};

{\color{\redC{}}
\draw [->, line width=\redT{}] (v1) -- (v2);
\draw [->, line width=\redT{}] (v2) to [out=220, in=50] (v3);
\draw [->, line width=\redT{}] (v3) to [out=105, in=255] (v1);
\draw [->, line width=\redT{}] (v2) to [out=325, in=35] (v4);
\draw [->, line width=\redT{}] (v4) to [out=115, in=245] (v2);
}

{\color{\blueC{}}
\draw [->, line width=\blueT{}] (v3) -- (v4);
\draw [->, line width=\blueT{}] (v4) -- (v1);
\draw [->, line width=\blueT{}] (v1) to [out=285, in=75] (v3);
\draw [->, line width=\blueT{}] (v2) to [out=295, in=65] (v4);
\draw [->, line width=\blueT{}] (v4) to [out=90, in=270] (v2);
}

\node [scale=0.8] at (3.5,-0.4) {$\mu{}(v_2,v_4)=2$};
\node [scale=0.8] at (3.5,-1.4) {$\mu{}(v_4,v_2)=2$};
\end{tikzpicture}\hspace{0.7cm}
\begin{tikzpicture}[scale=0.3]
 \node (v1) at (1.0,6.0) [vertexX] {$v_1$};
 \node (v2) at (6.0,6.0) [vertexX] {$v_2$};
 \node (v3) at (1.0,1.0) [vertexX] {$v_3$};
 \node (v4) at (6.0,1.0) [vertexX] {$v_4$};

{\color{\redC{}}
\draw [->, line width=\redT{}] (v1) to [out=35, in=145]  (v2);
\draw [->, line width=\redT{}] (v2) to [out=210, in=60] (v3);
\draw [->, line width=\redT{}] (v3) -- (v4);
\draw [->, line width=\redT{}] (v4) -- (v1);
}

{\color{\blueC{}}
\draw [->, line width=\blueT{}] (v1) -- (v2);
\draw [->, line width=\blueT{}] (v2) to [out=240, in=30] (v3);
\draw [->, line width=\blueT{}] (v1) --(v3);
\draw [->, line width=\blueT{}] (v3) to [out=130, in=230] (v1);
\draw [->, line width=\blueT{}] (v2) to [out=310, in=50] (v4);
\draw [->, line width=\blueT{}] (v4) -- (v2);
}

\node [scale=0.8] at (3.5,-0.4) {$\mu{}(v_1,v_2)=2$};
\node [scale=0.8] at (3.5,-1.4) {$\mu{}(v_2,v_3)=2$};
\end{tikzpicture}\hspace{0.7cm}
\begin{tikzpicture}[scale=0.3]
 \node (v1) at (1.0,6.0) [vertexX] {$v_1$};
 \node (v2) at (6.0,6.0) [vertexX] {$v_2$};
 \node (v3) at (1.0,1.0) [vertexX] {$v_3$};
 \node (v4) at (6.0,1.0) [vertexX] {$v_4$};

{\color{\redC{}}
\draw [->, line width=\redT{}] (v1) to [out=35, in=145]  (v2);
\draw [->, line width=\redT{}] (v2) -- (v3);
\draw [->, line width=\redT{}] (v3) to [out=340, in=200]  (v4);
\draw [->, line width=\redT{}] (v3) to [out=130, in=230] (v1);
\draw [->, line width=\redT{}] (v4) -- (v2);
}

{\color{\blueC{}}
\draw [->, line width=\blueT{}] (v1) -- (v2);
\draw [->, line width=\blueT{}] (v3) to [out=10, in=170]  (v4);
\draw [->, line width=\blueT{}] (v4) -- (v1);
\draw [->, line width=\blueT{}] (v1) --(v3);
\draw [->, line width=\blueT{}] (v2) to [out=310, in=50] (v4);
}

\node [scale=0.8] at (3.5,-0.4) {$\mu{}(v_1,v_2)=2$};
\node [scale=0.8] at (3.5,-1.4) {$\mu{}(v_3,v_4)=2$};
\end{tikzpicture}\hspace{0.7cm}
\begin{tikzpicture}[scale=0.3]
 \node (v1) at (1.0,6.0) [vertexX] {$v_1$};
 \node (v2) at (6.0,6.0) [vertexX] {$v_2$};
 \node (v3) at (1.0,1.0) [vertexX] {$v_3$};
 \node (v4) at (6.0,1.0) [vertexX] {$v_4$};

{\color{\redC{}}
\draw [->, line width=\redT{}] (v1) to [out=35, in=145]  (v2);
\draw [->, line width=\redT{}] (v4) -- (v1);
\draw [->, line width=\redT{}] (v1) to [out=280, in=80] (v3);
\draw [->, line width=\redT{}] (v3) to [out=140, in=220] (v1);
\draw [->, line width=\redT{}] (v2) to [out=310, in=50] (v4);
}

{\color{\blueC{}}
\draw [->, line width=\blueT{}] (v1) -- (v2);
\draw [->, line width=\blueT{}] (v2) -- (v3);
\draw [->, line width=\blueT{}] (v3) -- (v4);
\draw [->, line width=\blueT{}] (v3) to [out=110, in=250] (v1);
\draw [->, line width=\blueT{}] (v4) -- (v2);
}

\node [scale=0.8] at (3.5,-0.4) {$\mu{}(v_1,v_2)=2$};
\node [scale=0.8] at (3.5,-1.4) {$\mu{}(v_3,v_1)=2$};
\end{tikzpicture}
\caption{The \good{} decompositions given in the proof of Lemma~\ref{S4lemma}.}\label{figLem31}
\end{center}
\end{figure}

\begin{lem}\label{S4lemma}
Let $D$ be a directed multigraph with no arcs of multiplicity more than 2 and let $D$ contain $S_4$ as a spanning subdigraph. Then $D$ has no \good{} decomposition if and only if
$D$ is isomorphic to one of the following {\bf  exceptional digraphs}:
\begin{itemize}
\item $S_4$.
\item A directed multigraph obtained from $S_4$ by adding a copy of an arc in $S_4$ (isomorphic to $S_{4,1}$ or $S_{4,2}$, see Figure \ref{S4likefig}).
\item A directed multigraph obtained from $S_4$ by adding a copy of one arc in each of the two 2-cycles of $S_4$ (isomorphic to $S_{4,3}$, see Figure \ref{S4likefig}). 
\end{itemize}
\end{lem}
\begin{pf}
Observe that the Hamilton cycle $v_1v_2v_3v_4v_1$ is the only Hamilton cycle in $S_4$ and $|A(S_4)|=8$. 
Let $D' \in \{S_4, S_{4,1}, S_{4,2} , S_{4,3}\}$ be arbitrary and for the sake of contradiction assume that 
$D'$ has a \good{} decomposition consisting of strong subdigraphs $D_1,D_2$ with arc sets $A_1$ and $A_2$.

First consider the case when $|A_1|=4$, which implies that $A_1$ contains the arcs of the unique \GG{(up to copies of the same arc)} Hamilton cycle $v_1v_2v_3v_4v_1$.
However $D' - \{v_1v_2, v_2v_3, v_3v_4, v_4v_1\}$ is not strong for any digraph in $\{S_4, S_{4,1}, S_{4,2} , S_{4,3}\}$, implying that $|A_1| \geq 5$. Analogously $|A_2| \geq 5$, which implies that $D' = S_{4,3}$ and $|A_1|=|A_2|=5$.

In $D' = S_{4,3}$ we note that $d^+(v_4)=2$, so we may without loss of generality assume that 
$v_4v_2 \in A_1$ and $v_4 v_1 \in A_2$ and obtain the following:

\begin{itemize}
\item  $d^-(v_2)=2$  implies that $v_1 v_2 \in A_2$ (as $v_4v_2 \in A_1$). 
\item  $d^+(v_1)=2$  implies that $v_1 v_3 \in A_1$ (as $v_1v_2 \in A_2$). 
\item  $d^-(v_3)=2$  implies that $v_2 v_3 \in A_2$ (as $v_1v_3 \in A_1$).
\end{itemize}

Thus, all arcs from $\{v_2,v_4\}$ to $\{v_1,v_3\}$ belong to $A_2$, a contradiction \GG{with the assumption that $D_1$ is strong}.

Assume now that  $D$ is not isomorphic to any directed multigraph described in the statement of the lemma.
We will show that $D$ has a \good{} decomposition with subdigraphs with disjoint arc sets $A_1$ and $A_2$. We have four cases, which cover all possibilities subject to isomorphism.
It is not hard to check that the subdigraphs induced by both $A_1$ and $A_2$ given below are strong (see Figure~\ref{figLem31}).
\begin{description}
\item[Case 1:]   \jbj{$\mu{}(v_2,v_4)=\mu{}(v_4,v_2)=2$}. Then let $A_1=\{v_1v_2,v_2v_3,v_3v_1,v_2v_4,v_4v_2\}$ 
and $A_2=\{v_1v_3,v_3v_4,v_4v_1,v_2v_4,v_4v_2\}.$
\item[Case 2:] \jbj{$\mu{}(v_1,v_2)=\mu{}(v_2,v_3)=2$.} Then let $A_1$ contain the arcs of the Hamilton cycle of $S_4$ and $A_2$ the rest of the arcs of $D.$
\item[Case 3:] \jbj{$\mu{}(v_1,v_2)=\mu{}(v_3,v_4)=2$}. Then let  $A_1=\{v_1v_2,v_2v_3,v_3v_4,v_4v_2,v_3v_1\}$ and $A_2$ the rest of the arcs of $D.$
\item[Case 4:] \jbj{$\mu{}(v_1,v_2)=\mu{}(v_3,v_1)=2$}. Then let  $A_1=\{v_1v_2,v_2v_4,v_4v_1,v_1v_3,v_3v_1\}$ and $A_2$ the rest of the arcs of $D.$
\end{description}
It is not hard to check that up to isomorphism  all the exception digraphs are depicted in Figures \ref{S4fig} and \ref{S4likefig}.
\end{pf}

The following theorem was used in \cite{bangC24} to prove Theorem  \ref{thm03}. We will use it to prove Theorem \ref{sdmulti}.
Note that while in \cite{bangC24} Theorem \ref{ksubgraphs} was stated only for semicomplete digraphs, its proof in \cite{bangC24} shows that it holds 
also for semicomplete directed multigraphs.

\begin{thm}\label{ksubgraphs} \cite{bangC24}
Let $k\ge 1$ and let $D=(V,A)$ be a $k$-arc-strong semicomplete directed multigraph
such that there a set $S \subset V,$ with $2 \le |S| \le |V|-2$ and $|(S,V-S)_D| = k.$
Then there exist $k$ arc-disjoint strong spanning subgraphs of $D$ except if $D=S_4.$
\end{thm}

Now we are ready to prove the main result of this section.

\begin{thm}\label{sdmulti}
A 2-arc-strong semicomplete directed multigraph $D=(V,A)$ has a \good{} decomposition
if and only if it is not isomorphic to 
one of the \jbj{exceptional} digraphs depicted in Figures \ref{S4fig} and \ref{S4likefig}.
Furthermore, a \good{} decomposition of $D$ can be obtained in polynomial time when it exists.
\end{thm}
\begin{pf}
  We are going to prove the first part of the statement by induction over $n=|V|$ and then over the number of double arcs. The second part of the statement then follows as our proof is constructive. If there are no double arcs, then the claim follows from Theorem \ref{thm03}. 
If $n=2$ then $D$ is a directed multigraph consisting of two vertices $u,v$ \jbj{with $\mu{}(u,v)=\mu{}(v,u)=2$}.  Clearly this has a \good{} decomposition. 

Let $D$ be a semicomplete directed multigraph on at least 3 vertices with a double arc $uv$. If we can delete one copy of $uv$ and still have a 2-arc-strong semicomplete directed multigraph $D'$, then the claim follows by the induction hypothesis, so we may assume that $D'$ is not 2-arc-strong. Hence there is a partition $(X,V-X)$ of $V$ with $u\in X$ and $v\in V-X$ so that the two copies of $uv$ are the only arcs from $X$ to $V-X$. If $\min\{|X|,|V-X|\}\geq 2$, then it follows from Theorem \ref{ksubgraphs} that $D$ has a \good{} decomposition. Hence, we may assume w.l.o.g. that $X=\{u\}$.


Let $D^*$ be the digraph obtained from $D$ by contracting $\{u,v\}$ into one vertex, say $w$ (that is remove $\{u,v\}$ and
add $w$ such that for all $x \in V(D)\setminus \{u,v\}$ we have $\mu{}_{D^*}(w,x)= \mu{}_D(u,x) + \mu{}_D(v,x)$ and
$\mu{}_{D^*}(x,w)= \mu{}_D(x,u) + \mu{}_D(x,v)$).
Clearly $D^*$ is 2-arc-strong since any cut $(X,V-X)$ in $D^*$ gives a cut in $D$ (by replacing $w$ by $\{u,v\}$) with equaly many arcs
across. 
By the induction hypothesis, $D^*$ has a \good{} decomposition unless it has four vertices and is one of the exceptions from Lemma \ref{S4lemma}. However this is not the case as no vertex in these exceptions is an out-neighbour of all the other three vertices \jbj{(note that $\mu{}(x,w)\geq 1$ for all $w\in V(D^*)-w$ as $u$ is dominated by $V(D)-\{u,v\}$)}.
Hence $D^*$ has a \good{} decomposition $D_1,D_2$ and it remains to show that this can be modified to a \good{} decomposition of $D$.

Start by \jbj{replacing $w$ by $u,v$ in each of $D_1,D_2$ and then add a copy of $uv$ to each of the new versions of $D_1$ and $D_2$} and for the $\mu{}_{D^*}(x,w)$ arcs from every $x \in V\setminus \{u,v\}$
into $w$ in $D^*$ let 
$\mu{}_D(x,u)$ of these go to $u$ and $\mu{}_D(x,v)$ of these go to $v$ (keeping them in the same $D_i$ as they were before). 
Note that all arcs out of $w$ in $D^*$ correspond to arcs out of $v$ in $D$.  If we can do the above procedure such that 
$u$ receives an arc into it in both $D_1$ and $D_2$, then we obtain a \good{} decomposition of $D$. Furthermore this is always possible
if there exist two vertices $z_1,z_2$ so that $z_i$ is an in-neighbour of $v$ in $D_i$ (in $D^*$), $i=1,2$ and either $z_1\neq z_2$ or $z_1=z_2$ and this vertex has a double arc to $u$ in $D$. So we may assume that this is not the case, which implies that $n=3$ and $V=\{u,v,z\}$.
 By the arguments above and the fact that $D$ is 2-arc-strong we get that $vz$ is a double arc and there is at least one arc 
from $v$ to $u$ since $zu$ is not a double arc. Now we get the desired
\good{} decomposition by taking the two sets of arcs $\{uv,vz,zu\}$ and $\{uv,vu,vz,zv\}$.

\end{pf}

\section{Semicomplete compositions containing a cut-vertex}\label{sec:cutvertex}

In this section we will prove the following lemma, which will turn out to be very useful in the proofs below, and is of
interest in its own right.

\begin{lem}\label{cutvertex}
Let $D=T[H_1,\dots , H_t]$ be 2-arc-strong ($t \geq 2$),  where $T$ is a semicomplete digraph and every $H_i$ is an arbitrary digraph.
If $D$ contains a cut-vertex then $D$ has a \good{} decomposition.
\end{lem}

\begin{pf} 
Let $u$ be a cut-vertex in $D$ and let $D'=D - u$. Let $(X,Y)$ be a cut in $D'$ such that there is no arc from $Y = V(D') - X$ to $X$
in $D'$. Let $D^* = \induce{D}{X \cup \{u\}}$  and let $D^{**} = \induce{D}{Y \cup \{u\}}$.  Note that
$|V(D^*)| \geq 2$ and $|V(D^{**})| \geq 2$. We now prove the following claims.

\2

{\bf  Claim 1.} We may assume that $D'$ contains vertices from more than one $H_i$.

\2

{\bf  Proof of Claim 1:} For the sake of contradiction, assume that $V(D') \subseteq V(H_1)$. 
Since $D=T[H_1,\dots , H_t]$ and $t \geq 2$, this implies that $t=2$, $T$ is a $2$-cycle, $V(H_2) = \{u\}$ \GG{and for every $v\in V(H_1)$ we have $vu,uv\in A(D)$}.
Let $Q_1,Q_2,\ldots,Q_l$ be strong components in $D'$, such that there is no arc from $Q_i$ to $Q_j$ when $i>j$.
We will now construct a \good{} decomposition, $(G_1,G_2)$ of $D$ as follows.

For all $Q_i$ with $|V(Q_i)| \geq 2$, do the following.  Let $x_iy_i$ be any arc in $Q_i$.
Add the arcs $(A(Q_i) \setminus \{x_iy_i\}) \cup \{x_i u, u y_i\}$ to $G_1$ and add all arcs $\{u x_i, x_i y_i, y_i u\}$ and all arcs
$\{uw, wu\}$ for all $w \in V(Q_i)\setminus \{x_i,y_i\}$ to $G_2$. Note that this implies that $\induce{G_a}{V(Q_i) \cup \{u\}}$ is
strong for $a=1,2$.

Now add all arcs between different $Q_i$'s to $G_1$. Furthermore for all $Q_i$ with $|V(Q_i)| = 1$ assume that $V(Q_i)=\{x_i\}$ and add
$\{x_i u , u x_i\}$ to $G_2$.
As $d^+(x_i) \geq 2$ and $d^-(x_i) \geq 2$ in $D$  we note that $2 \leq i \leq l-1$ and $x_i$ has an arc into it from a $Q_j$ with
$j <i$ and an arc out of it to a $Q_k$ with $k>i$ and these arcs belong to $G_1$. It follows that $x_i$ therefore
belongs to a path from a $Q_a$ to a $Q_b$ in $G_1$, where $a<b$ and $|V(Q_a)| \geq 2$ and $|V(Q_b)| \geq 2$. Therefore
$(G_1,G_2)$ is a \good{} decomposition in $D$, which completes the proof of Claim~1.

\2

{\bf  Claim 2.} We may assume that $X \cap V(H_i) = \emptyset$ or $Y \cap V(H_i) = \emptyset$ for all $i=1,2,\ldots,t$.

\2

{\bf  Proof of Claim 2:} For the sake of contradiction, assume \jbj{w.l.o.g.} that $X \cap V(H_1) \not= \emptyset$ and $Y \cap V(H_1) \not= \emptyset$.
By Claim~1 either $X$ or $Y$ contains a vertex not in $H_1$.  Assume without loss of generality that $Y \setminus V(H_1) \not= \emptyset$.
Let $X' = X \cup V(H_1) \setminus \{u\}$ and let $Y' = Y \setminus (V(H_1) \cup \{u\})$. Note that 
$(X',Y')$ is a cut in $D'$ with no arc from $Y'$ to $X'$ as otherwise there are arcs from $Y'$ to $H_1$, \jbj{contradicting that there is no arc from $Y$ to $X.$}
The process \GG{of moving from $(X,Y)$ to $(X',Y')$} decreased the number of $H_i$ with vertices in both $X$ and $Y$, so continuing this process we will obtain that 
$X \cap V(H_i) = \emptyset$ or $Y \cap V(H_i) = \emptyset$ for all $i=1,2,\ldots,t$.  This completes the proof of Claim~2.

\2

{\bf  Claim 3.} There exists two arc-disjoint out-branchings in $D^*$ both rooted at $u$
and there exists two arc-disjoint in-branchings in $D^{**}$ both rooted at $u$.

\2

{\bf  Proof of Claim 3:} Let $w \in X$ be arbitrary. 
As $D$ is $2$-arc-strong there are $2$ arc-disjoint paths, $P_1$ and $P_2$, from $u$ to $w$ in $D$. 
We note that $V(P_1) \subseteq V(D^*)$ and $V(P_2) \subseteq V(D^*)$, as any path from a vertex not in $X$ to $w$ goes through $u$.
Therefore there exists two arc-disjoint paths from $u$ to $w$ in $D^*$.
By Edmonds' branching theorem, Theorem \ref{Edmthm}, there exists two arc-disjoint out-branchings in $D^*$ both rooted at $u$.

Analogously if  $w \in Y$ then there exist two arc-disjoint paths from $w$ to $u$ in $D$ and therefore also in $D^{**}$.
Again, by Theorem~\ref{Edmthm}, there exists two arc-disjoint in-branchings in $D^{**}$ both rooted at $u$.
This completes the proof of Claim~3.

\2

{\bf Definition of $O_1^*$, $O_2^*$, $I_1^{**}$ and $I_2^{**}$:} 
Let $O_1^*$ and $O_2^*$ be two arc-disjoint out-branchings in $D^*$ rooted at $u$ and
let $I_1^{**}$ and $I_2^{**}$ be two arc-disjoint in-branchings in $D^{**}$ rooted at $u$ found in Claim~3.

\2

{\bf  Claim 4.} Each branching $O_1^*$, $O_2^*$, $I_1^{**}$ and $I_2^{**}$ contains a vertex that is neither a leaf nor the root $u$.

Denote these vertices by $o_1^*$, $o_2^*$, $i_1^{**}$ and $i_2^{**}$, respectively (see Figure~\ref{o1o2}).

\2

{\bf  Proof of Claim 4:} Consider $O_1^*$. 
As $|V(D^*)| \geq 2$ we note that there is an arc $u v \in O_2^*$. This implies that $u v \not\in O_1^*$ 
(as we do not have parallel arcs in $D$).
However some  arc, say $o_1^* v$, enters $v$ in $O_1^*$.
Now $o_1^*$ is not the root of $O_1^*$ and is also not a leaf of $O_1^*$. 
The cases of $O_2^*$, $I_1^{**}$ and $I_2^{**}$ can be proved analogously, which completes the proof of Claim~4.

\2

{\bf  Definition of $G_1$ and $G_2$:} We now define an arc decomposition of $D$ as follows 
(see Figures~\ref{o1o2} and \ref{G1G2} for an illustration).
Let $G_1$ contain all arcs of $O_1^*$ and $I_1^{**}$ and let $G_2$ contain all arcs of $O_2^*$ and $I_2^{**}$.
Note that all arcs between $o_1^*$ and $V(D) \setminus V(D^*)$ exist (by Claim~2 and the fact that $T$ is semicomplete) 
and go out of $o_1^*$. 
Add all arcs from $o_1^*$ to $V(D) \setminus (V(D^*) \cup \{i_2^{**}\})$ to $G_2$.
Analogously add all arcs from $o_2^*$ to $V(D) \setminus (V(D^*)\cup \{i_1^{**}\})$ to $G_1$.
Also, add all arcs from $V(D) \setminus (V(D^{**} \cup \{o_2^{*}\})$ to $i_1^{**}$ to $G_2$ and
add all arcs from $V(D) \setminus (V(D^{**} \cup \{o_1^{*}\})$ to $i_2^{**}$ to $G_1$.
Any remaining arcs from $D$ which have not been added to $G_1$ or $G_2$ yet can be added arbitrarily.
This completes the definition of $G_1$ and $G_2$.

\2

\begin{figure} 
\tikzstyle{vertexX}=[circle,draw, top color=gray!5, bottom color=gray!30, minimum size=15pt, scale=0.5, inner sep=1.0pt]
\begin{tikzpicture}[scale=0.25]
 \node (uj1) at (5.0,17.0) [vertexX] {$u$};
 \node (x1) at (1.0,13.0) [vertexX] {};
 \node (x2) at (1.0,9.0) [vertexX] {};
 \node (x3) at (1.0,5.0) [vertexX] {};
 \node (x4) at (1.0,1.0) [vertexX] {$o_1^*$};
{\color{\redC{}}
\draw [->, line width=\redT{}] (x4) -- (x3);
\draw [->, line width=\redT{}] (uj1) to [out=200, in=150] (x2);
\draw [->, line width=\redT{}] (uj1) to [out=160, in=150] (x4);
\draw [->, line width=\redT{}] (x3) to [out=120, in=240] (x1);
}
\node at (3,-0.5) {$O_1^*$};
\end{tikzpicture} \hspace{0.5cm} 
\begin{tikzpicture}[scale=0.25]
 \node (uj1) at (5.0,17.0) [vertexX] {$u$};
 \node (x1) at (1.0,13.0) [vertexX] {};
 \node (x2) at (1.0,9.0) [vertexX] {$o_2^*$};
 \node (x3) at (1.0,5.0) [vertexX] {};
 \node (x4) at (1.0,1.0) [vertexX] {};
{\color{\blueC{}}
\draw [->, line width=\blueT{}] (uj1) -- (x1);
\draw [->, line width=\blueT{}] (x1) -- (x2);
\draw [->, line width=\blueT{}] (x2) -- (x3);
\draw [->, line width=\blueT{}] (x2) to [out=240, in=120] (x4);
}
\node at (3,-0.5) {$O_2^*$};
\end{tikzpicture} \hspace{1.0cm} 
\begin{tikzpicture}[scale=0.25]
 \node (uj1) at (5.0,17.0) [vertexX] {$u$};
 \node (y1) at (9.0,13.0) [vertexX] {};
 \node (y2) at (9.0,9.0) [vertexX] {};
 \node (y3) at (9.0,5.0) [vertexX] {$i_1^{**}$};
 \node (y4) at (9.0,1.0) [vertexX] {};
{\color{\redC{}}
\draw [->, line width=\redT{}] (y1) -- (uj1);
\draw [->, line width=\redT{}] (y3) to [out=30, in=0] (uj1);
\draw [->, line width=\redT{}] (y2) -- (y1);
\draw [->, line width=\redT{}] (y4) -- (y3);
}
\node at (7,-0.5) {$I_1^{**}$};
\end{tikzpicture} \hspace{0.5cm} 
\begin{tikzpicture}[scale=0.25]
 \node (uj1) at (5.0,17.0) [vertexX] {$u$};
 \node (y1) at (9.0,13.0) [vertexX] {};
 \node (y2) at (9.0,9.0) [vertexX] {$i_2^{**}$};
 \node (y3) at (9.0,5.0) [vertexX] {};
 \node (y4) at (9.0,1.0) [vertexX] {};
{\color{\blueC{}}
\draw [->, line width=\blueT{}] (y2) to [out=30, in=340] (uj1);
\draw [->, line width=\blueT{}] (y4) to [out=30, in=20] (uj1);
\draw [->, line width=\blueT{}] (y1) to [out=300, in=60] (y3);
\draw [->, line width=\blueT{}] (y3) -- (y2);
}
\node at (7,-0.5) {$I_2^{**}$};
\end{tikzpicture} 
\caption{An example of $O_1^*$, $O_2^*$, $I_1^{**}$ and $I_2^{**}$ and $o_1^*$, $o_2^*$, $i_1^{**}$ and $i_2^{**}$.}
\label{o1o2}
\end{figure}
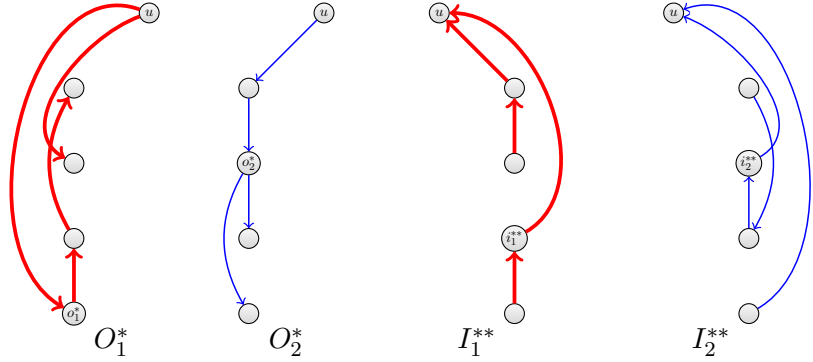

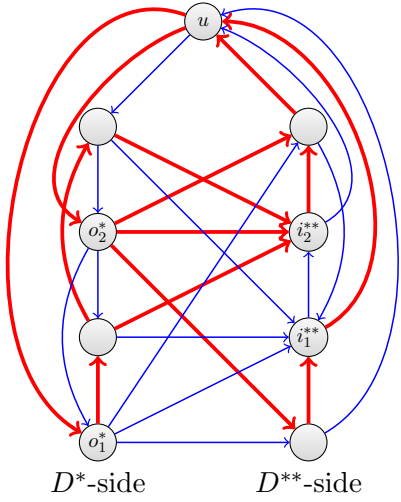
\begin{figure}[h!]
\tikzstyle{vertexX}=[circle,draw, top color=gray!5, bottom color=gray!30, minimum size=20pt, scale=0.7, inner sep=1.5pt]
\tikzstyle{vertexY}=[circle,draw, top color=gray!5, bottom color=gray!30, minimum size=20pt, scale=0.7, inner sep=1.5pt]
\begin{center}
\begin{tikzpicture}[scale=0.35]
 \node (uj1) at (5.0,17.0) [vertexX] {$u$};
 \node (x1) at (1.0,13.0) [vertexX] {};
 \node (x2) at (1.0,9.0) [vertexX] {$o_2^*$};
 \node (x3) at (1.0,5.0) [vertexX] {};
 \node (x4) at (1.0,1.0) [vertexX] {$o_1^*$};
 \node (y1) at (9.0,13.0) [vertexX] {};
 \node (y2) at (9.0,9.0) [vertexX] {$i_2^{**}$};
 \node (y3) at (9.0,5.0) [vertexX] {$i_1^{**}$};
 \node (y4) at (9.0,1.0) [vertexX] {};

{\color{\redC{}}
\draw [->, line width=\redT{}] (x4) -- (x3);
\draw [->, line width=\redT{}] (uj1) to [out=200, in=150] (x2);
\draw [->, line width=\redT{}] (uj1) to [out=160, in=150] (x4);
\draw [->, line width=\redT{}] (x3) to [out=120, in=240] (x1);
\draw [->, line width=\redT{}] (y1) -- (uj1);
\draw [->, line width=\redT{}] (y3) to [out=30, in=0] (uj1);
\draw [->, line width=\redT{}] (y2) -- (y1);
\draw [->, line width=\redT{}] (y4) -- (y3);
\draw [->, line width=\redT{}] (x2) -- (y1);
\draw [->, line width=\redT{}] (x2) -- (y2);
\draw [->, line width=\redT{}] (x2) -- (y4);
\draw [->, line width=\redT{}] (x1) -- (y2);
\draw [->, line width=\redT{}] (x3) -- (y2);
}

{\color{\blueC{}}
\draw [->, line width=\blueT{}] (uj1) -- (x1);
\draw [->, line width=\blueT{}] (x1) -- (x2);
\draw [->, line width=\blueT{}] (x2) -- (x3);
\draw [->, line width=\blueT{}] (x2) to [out=240, in=120] (x4);
\draw [->, line width=\blueT{}] (y2) to [out=30, in=340] (uj1);
\draw [->, line width=\blueT{}] (y4) to [out=30, in=20] (uj1);
\draw [->, line width=\blueT{}] (y1) to [out=300, in=60] (y3);
\draw [->, line width=\blueT{}] (y3) -- (y2);
\draw [->, line width=\blueT{}] (x4) -- (y1);
\draw [->, line width=\blueT{}] (x4) -- (y3);
\draw [->, line width=\blueT{}] (x4) -- (y4);
\draw [->, line width=\blueT{}] (x1) -- (y3);
\draw [->, line width=\blueT{}] (x3) -- (y3);
}
\node at (1,-0.5) {$D^*$-side};
\node at (9,-0.5) {$D^{**}$-side};
\end{tikzpicture}
\end{center}
\caption{An illustration of $G_1$ and $G_2$ obtained from $O_1^*$, $O_2^*$, $I_1^{**}$ and $I_2^{**}$ seen in Figure~\ref{o1o2}. 
The thick arcs give us $G_1$ and the thin arcs $G_2$. Note that both $G_1$ and $G_2$ induce strong spanning subdigraphs.}
\label{G1G2} 
\end{figure}

{\bf  Claim 5:} $(G_1,G_2)$ is a \good{} decomposition of $D$. 

\2

{\bf  Proof of Claim 5:} First let $v \in X$ be arbitrary. 
We will now show that there exists a $(v,u)$-path in $G_1$.
As $O_1^* \subseteq G_1$, there is a path, $P_1$, from $v$ to a leaf $l^*$ in $O_1^*$. 
By construction the arc $l^* i_2^{**}$ belongs to $G_1$ 
(as $l^* \not= o_1^*$, since $l^*$ is a leaf in $O_1^*$ and $o_1^*$ is not a leaf). 
As $I_1^{**} \subseteq G_1$, there is a path, $P_2$, from $i_2^{**}$ to $u$ in $I_1^{**}$. The path $P_1 P_2$ is now the desired
$(v,u)$-path in $G_1$.

Analogously we can show that there exists a $(v,u)$-path in $G_2$.
Furthermore as $O_1^*$ and $O_2^*$ are out-branchings in $G_1$ and $G_2$, respectively we can also find a $(u,v)$-path in both $G_1$ and $G_2$.

Let $w \in Y$ be arbitrary.  Analogously we can find a $(u,w)$-path in both $G_1$ and $G_2$, by considering a path from
a leaf $l_r^{**}$ in $G_r$ ($r \in \{1,2\}$) such that there exists a $(l_r^{**},w)$-path in $I_r^{**}$ and noting that 
$o_{3-r}^* l_r^{**}$ is an arc in $G_r$ and there exists a $(u,o_{3-r}^*)$-path in $O_r^*$.
As $I_1^{**}$ and $I_2^{**}$ are in-branchings in $G_1$ and $G_2$, respectively we can also find a $(w,u)$-path in both $G_1$ and $G_2$.

This implies that every vertex in $D'$ has a path to $u$ and a path from $u$ in $G_1$ and in $G_2$\jbj{,  showing that }
$G_1$ and $G_2$ is a \good{} decomposition, thereby proving Claim 5 and the lemma.
\end{pf}

\section{Main Results}\label{sec:main}



Our main technical result is the following theorem. Recall that by Remark \ref{viewrem}, $\overrightarrow{C}_3[\overline{K}_2,\overline{K}_2,\overrightarrow{P}_2]=T^s_4[\overline{K}_2,\overline{K}_2,\overline{K}_1,\overline{K}_1],$
where $T^s_4$ is the unique strong tournament on four vertices, see Figure 2.

\GG{\begin{thm}\label{mainX}
  Let $D=T[\overline{K}_{n_1},\ldots{},\overline{K}_{n_t}]$ be an extended semicomplete digraph where $n_i\leq 2$ for  $i\in [t]$.
  If $D$ is $2$-arc-strong,
then $D$ has a \good{} decomposition if and only if $D$ is not isomorphic to to one of the following three digraphs: $S_4,$
$\overrightarrow{C}_3[\overline{K}_2,\overline{K}_2,\overline{K}_2]$, $\overrightarrow{C}_3[\overline{K}_2,\overline{K}_2,\overrightarrow{P}_2]=T^s_4[\overline{K}_2,\overline{K}_2,\overline{K}_1,\overline{K}_1].$
\end{thm}

}

\GG{Before proving this theorem in the next section, we use it to prove Theorem \ref{mainY}, which is the special case of our main result, Theorem \ref{mainT}, for 
extended semicomplete digraphs. Theorem \ref{mainY} and Lemma \ref{lemGeneral1} will directly imply Theorem \ref{mainT}.

\begin{thm} \label{mainY}
Let $D=T[\overline{K}_{n_1},\ldots{},\overline{K}_{n_t}]$ be a 2-arc-strong extended semicomplete digraph. Then $D$   has a \good{} decomposition if and only if $D$ is not isomorphic 
to one of the following four digraphs: $S_4,$
$\overrightarrow{C}_3[\overline{K}_2,\overline{K}_2,\overline{K}_2]$, $\overrightarrow{C}_3[\overline{K}_2,\overline{K}_2,\overrightarrow{P}_2],$
$\overrightarrow{C}_3[\overline{K}_2,\overline{K}_2,\overline{K}_3].$
\end{thm}
}
\begin{pf}
\GG{Let ${\cal C}=\{\overrightarrow{C}_3[\overline{K}_2,\overline{K}_2,\overline{K}_2], \overrightarrow{C}_3[\overline{K}_2,\overline{K}_2,\overrightarrow{P}_2],
\overrightarrow{C}_3[\overline{K}_2,\overline{K}_2,\overline{K}_3]\}.$}
We will prove the theorem by induction over $|V(D)|$. If $|V(D)| \leq 3$, then the theorem clearly holds, \jbj{so the base case holds}.
  If $n_i \leq 2$ for all $i=1,2,\ldots,t$ then we are done by Theorem~\ref{mainX}, so \GG{we may 
assume that $n_j \geq 3$ for some $j$}. Let $D' = D - u_{j,n_j}$.  \GG{Since $n_j-1\ge 2$, $D'\neq S_4.$}

It follows from Lemma \ref{extended2AS} that $D'$ is $2$-arc-strong
and hence it fulfils the statement of the theorem by induction.
If $D'\not\in {\cal C}$, then, by induction, it has a \good{} decomposition and hence $D$ also has a \good{} decomposition by  Lemma \ref{lemo}. Hence we may assume that $D'\in {\cal C}$ \GG{and consider the corresponding three cases.}

\2

{\bf  Case 1: $D' = \overrightarrow{C}_3[\overline{K}_2,\overline{K}_2,\overline{K}_2]$.} This implies that 
$D = \overrightarrow{C}_3[\overline{K}_2,\overline{K}_2,\overline{K}_3]$, and therefore $D \in {\cal C}$, which completes this case.

\2

{\bf  Case 2: $D' = \overrightarrow{C}_3[\overline{K}_2,\overline{K}_2,\overline{K}_3]$.} In this case
$D = \overrightarrow{C}_3[\overline{K}_2,\overline{K}_3,\overline{K}_3]$ or $D = \overrightarrow{C}_3[\overline{K}_2,\overline{K}_2,\overline{K}_4]$.
These digraphs have \good{} decompositions by Theorem \ref{good-decompSemi}.

\2

{\bf  Case 3: $D' = \overrightarrow{C}_3[\overline{K}_2,\overline{K}_2,\overrightarrow{P}_2]$.} \GG{As in Case 2, $D$ has  a \good{} decomposition by Theorem \ref{good-decompSemi}.}

\2

This completes the proof of the theorem.  
\end{pf}

\GG{Now Theorem \ref{mainY} and the following lemma imply our main result, Theorem \ref{mainT}. }

\begin{lem}\label{lemGeneral1}
Let $D=T[H_1,\dots , H_t]$ be 2-arc-strong ($t \geq 2$),  where $T$ is a semicomplete digraph and every $H_i$ is an arbitrary digraph.
Then one of the following cases holds.
\begin{description}
\item[(a)]  $D$ \jbj{has} a \good{} decomposition.
\item[(b)]  $D$ is an extended semicomplete digraph.
\item[(c)]  \jbj{For every $i\in [t]$ and every arc $e$ of  $H_i$, $D-e$ is 2-arc-strong.}
\end{description}
\end{lem}
 \begin{pf} 
For the sake of contradiction assume that none of (a)-(c) hold and let $D=(V,A)$.
As $D$ is not an extended semicomplete digraph (otherwise (b) holds) there exists an arc  in some $H_i$ \jbj{and since (c) does not hold we can choose $i\in [t]$ and an arc $e=uv$ of $H_i$ such that the digraph 
$D' = D -e$ is strong, but not 2-arc-strong}.
Let $(X,V-X)$ be a cut in $D'$ such that there is only one arc, $xy$, from $X$ to $V-X$ and note that 
$u \in X$ and $v \in V-X$ as $D$ is 2-arc-strong. Note that either $x \not=u$ or $y \not= v$, and we assume without loss of generality 
that $y \not= v$. This implies that $v$  has no arc into it from $X$ \GG{in $D'$}. 

First consider the case when $X \setminus V(H_i) \not= \emptyset$. In this case let $Y = X \setminus V(H_i)$ and 
note that $(Y,V-Y)$ is a cut in $D$ with at most one arc from $Y$ to $V-Y$ (\jbj{the only possible arc is the arc $xy$ since if $z\in Y$ has an arc to a vertex in $V(H_i)\cap X$ then $zv$ is an arc from $X$ to $V-X$}), a contradiction
to $D$ being 2-arc-strong.  We may therefore assume that $X \setminus H_i = \emptyset$, which is equivalent to $X \subseteq V(H_i)$.

If $y \in V(H_i)$ then \GG{$N_{D'}^+(x) \subseteq V(H_i)$} which implies that \jbj{there is no arc leaving
  $V(H_i)$ in $D'$, contradicting that $D'$ is strong} (and $t \geq 2$).
Therefore $y \not\in V(H_i)$ and $N_{D'}^+(V(H_i)) = \{y\}$. If $V(D) \not= H_i \cup \{y\}$ then we note that
$y$ is a cut-vertex in \GG{$D'$}, separating $H_i$ from $V \setminus (H_i \cup \{y\})$ in \GG{$D'$}. In this case there is a
\good{} decomposition in \GG{$D'$} by Lemma~\ref{cutvertex}, a contradiction to (a) not holding \GG{as clearly $D$ has a
\good{} decomposition, too)}.  

We may therefore assume that $V(D) = V(H_i) \cup \{y\}$. \GG{
}
By Lemma~\ref{cutvertex} we may assume that $H_i$ is strongly connected, as $y$ otherwise would be a cut-vertex.
Let us consider the following pair $G_1,G_2$ of disjoint spanning subdigraphs of $D$. 
The arcs of $G_1$ are $(A(H_i) \setminus \{uv\}) \cup \{u y, y v\}$ and the
arcs of $G_2$ are $\{y u, u v, v y\}$ and all arcs
$\{yw, wy\}$ for all $w \in V(H_i)\setminus \{u,v\}$. Since both $G_1$ and $G_2$ are strong,
$D$ has a \good{} decomposition, contradicting the assumption that (a) does not hold, and thereby completing the proof.
 \end{pf}

Recall the statement of Theorem~\ref{mainT}.

\vspace{0.2cm}

\noindent{\bf Theorem~\ref{mainT}}
{\em Let $T$ be a strong semicomplete digraph on $t\ge 2$ vertices and let $H_1,\dots ,H_t$ be arbitrary digraphs.
Then $D=T[H_1,\dots ,H_t]$ has a \good{} decomposition if and only if $D$ is 2-arc-strong and is not isomorphic to one of the following four digraphs: $S_4,$
$\overrightarrow{C}_3[\overline{K}_2,\overline{K}_2,\overline{K}_2]$, $\overrightarrow{C}_3[\overline{K}_2,\overline{K}_2,\overrightarrow{P}_2],$
$\overrightarrow{C}_3[\overline{K}_2,\overline{K}_2,\overline{K}_3].$} 

\vspace{0.2cm}
\GG{
\begin{pf}
By Theorems \ref{thm03} and \ref{good-decompSemi}, $S_4,$
$\overrightarrow{C}_3[\overline{K}_2,\overline{K}_2,\overline{K}_2]$, $\overrightarrow{C}_3[\overline{K}_2,\overline{K}_2,\overrightarrow{P}_2],$
$\overrightarrow{C}_3[\overline{K}_2,\overline{K}_2,\overline{K}_3]$ have no \good{} decompositions. 
Suppose that $D$ satisfies the conditions of the theorem and yet has no \good{} decomposition. Then $D$ satisfies either Case (b) or (c) of Lemma \ref{lemGeneral1}.
However, Case (c) of Lemma \ref{lemGeneral1} can be reduced to Case (b). Thus,
Theorem \ref{mainY} implies Theorem \ref{mainT}.
\end{pf}
}

\section{Proof of Theorem \ref{mainX}}\label{sec:X}
Before giving the proof of Theorem~\ref{mainX} we will prove the following lemma, which is needed in the proof of
Theorem~\ref{mainX}.

\begin{lem}
  \label{endofcutarc}
  Let $D=T[H_1,\ldots{},H_t]$ be a 2-arc-strong extended semicomplete digraph which has no cut-vertex, $V(T)=\{u_1,\dots ,u_t\}$, $|V(H_i)|\leq 2$ for $i\in [t]$ and $V(H_r)=\{x,y\}$. Suppose $D'=D-y$ is not 2-arc strong. Then there exists an index $q\neq r$ such that one of the following holds,

\begin{description}
\item[(i)] $u_ru_q$ is a cut-arc of $T$ and $N^-(V(H_q))=V(H_r)$, or
\item[(ii)] $u_qu_r$ is a cut-arc of $T$ and $N^+(V(H_q))=V(H_r)$.
\end{description}
 \end{lem}

  \begin{pf} As $D'$ is strong (since $D$ contains no cut-vertex) but not 2-arc-strong, there is a proper subset $S$ of $V(D')$ such that
there is exactly one arc $uv$ from $S$ to $\bar{S}=V(D)-S$ in $D'$. Suppose first that $x\in S$. 
If   $x \neq u$ then $(S+y,\bar{S})$ is a vertex partition of $V(D)$ with only one arc from $S+y$ to $\bar{S}$, 
contradicting that $D$ is 2-arc-strong (here we used the fact that $x$ and $y$ have the same out-neighbours).
 Thus we must have $x=u$ which implies that $u_ru_q$ is a cut-arc of $T$, where $v \in V(H_q)$, 
as $uv$ is the only arc from $S'$ to $V(T)-S'$ 
where $S'\subset V(T)$ is the set of vertices in $T$ that we obtain by taking $u_j$ in $S'$ precisely when 
$V(H_j)\cap S\neq\emptyset$.

If  $|V(H_q)|=2$, then we may assume that $H_q=\{v,w\}$ for some $w\neq v$. In this case we must have $w\in S$ as $xw\in A(D)$. 
Since $v$ is not a cut-vertex we must have $\bar{S}=\{v\}$. Similarly if 
$|V(H_q)|=1$ then $\bar{S}=\{v\}$, as otherwise $v$ is a cut-vertex in $D$.
As $\bar{S}=\{v\}$, we note that $N^-(V(H_q))=V(H_r)$, implying that Part (i) of the lemma holds in this case. 

It is easy to see that case when $x\in\bar{S}$ leads to Part (ii) of the lemma.  \end{pf}

Recall the statement of Theorem~\ref{mainX}.

\vspace{0.2cm}

\GG{
\noindent{\bf Theorem~\ref{mainX}}
{\em  Let $D=T[\overline{K}_{n_1},\ldots{},\overline{K}_{n_t}]$ be an extended semicomplete digraph where $n_i\leq 2$ for  $i\in [t]$.
  If $D$ is $2$-arc-strong, then $D$ has a \good{} decomposition if and only if $D$ is not isomorphic to to one of the following three digraphs: $S_4,$
$\overrightarrow{C}_3[\overline{K}_2,\overline{K}_2,\overline{K}_2]$, $\overrightarrow{C}_3[\overline{K}_2,\overline{K}_2,\overrightarrow{P}_2]=T^s_4[\overline{K}_2,\overline{K}_2,\overline{K}_1,\overline{K}_1].$}
}

\vspace{0.2cm}

\begin{pf} 
Let ${\cal D}_2=\{S_4,\overrightarrow{C}_3[\overline{K}_2,\overline{K}_2,\overline{K}_2],\overrightarrow{C}_3[\overline{K}_2,\overline{K}_2,\overrightarrow{P}_2]\}$ and 
$D$ and $T$ defined as in the theorem and $V(T)=\{u_1,\dots ,u_t\}.$ \jbj{For all $i\in [t]$ let $H_i$ denote the $i$'th subdigraph in the decomposition, i.e. $H_i=\overline{K}_{n_i}$ 
and denote  the vertices of $H_i$ by} $u_{i,j_i},\ 1\le j_i\le n_i$. 
By the assumption in the theorem $n_i \in \{1,2\}$ for all $i\in [t]$.
We consider the following cases.

\2

{\bf  Case 1.} $|V(T)|\leq 2$.

\2

As $D$ is $2$-arc-strong we must have $|V(T)| = t = 2$, 
$A(T)=\{u_1u_2, u_2u_1\}$ and
$|H_1|=|H_2|=2$. Then $u_{1,1} u_{2,1} u_{1,2} u_{2,2} u_{1,1}$ and 
$u_{1,1} u_{2,2} u_{1,2} u_{2,1} u_{1,1}$ form arc-disjoint Hamilton cycles (see \AY{Figure~\ref{fig_x1}(a)}), thereby proving that $D$ has a \good{} decomposition.

\2

\begin{figure}[!htb] 
\tikzstyle{vertexX}=[circle,draw, top color=gray!5, bottom color=gray!30, minimum size=16pt, scale=0.6, inner sep=0.5pt]
\tikzstyle{vertexY}=[circle,draw, top color=gray!5, bottom color=gray!30, minimum size=27pt, scale=0.9, inner sep=2.5pt]
\begin{tikzpicture}[scale=0.45]
 \node (u11) at (1.0,6.0) [vertexX] {$u_{1,1}$};
 \node (u12) at (1.0,1.0) [vertexX] {$u_{1,2}$};
 \node (u21) at (6.0,6.0) [vertexX] {$u_{2,1}$};
 \node (u22) at (6.0,1.0) [vertexX] {$u_{2,2}$};

{\color{\blueC{}}
\draw [->, line width=\blueT{}] (u11) to [out=15, in=165] (u21);
\draw [->, line width=\blueT{}] (u21) to [out=210, in=60] (u12);
\draw [->, line width=\blueT{}] (u12) to [out=345, in=195] (u22);
\draw [->, line width=\blueT{}] (u22) to [out=150, in=300] (u11);
}

{\color{\redC{}}
\draw [->, line width=\redT{}] (u11) to [out=330, in=120] (u22);
\draw [->, line width=\redT{}] (u22) to [out=165, in=15] (u12);
\draw [->, line width=\redT{}] (u12) to [out=30, in=240] (u21);
\draw [->, line width=\redT{}] (u21) to [out=195, in=345] (u11);
}

\node at (3.5,-0.5) {$(a)$};
\end{tikzpicture} \hspace{0.85cm}
\begin{tikzpicture}[scale=0.45]
 \node (xx)  at (3.5,8.5) [vertexX] {$u_{1,1}$};
 \node (u11) at (1.0,6.0) [vertexX] {$u_{2,1}$};
 \node (u12) at (1.0,1.0) [vertexX] {$u_{2,2}$};
 \node (u21) at (6.0,6.0) [vertexX] {$u_{3,1}$};
 \node (u22) at (6.0,1.0) [vertexX] {$u_{3,2}$};

{\color{\blueC{}}
\draw [->, line width=\blueT{}] (u11) to [out=15, in=165] (u21);
\draw [->, line width=\blueT{}] (u21) to [out=210, in=60] (u12);
\draw [->, line width=\blueT{}] (u12) to [out=345, in=195] (u22);
\draw [->, line width=\blueT{}] (u22) to [out=150, in=300] (u11);
\draw [->, line width=\blueT{}] (xx) -- (u12);
\draw [->, line width=\blueT{}] (u22) -- (xx);
}

{\color{\redC{}}
\draw [->, line width=\redT{}] (u11) to [out=330, in=120] (u22);
\draw [->, line width=\redT{}] (u22) to [out=165, in=15] (u12);
\draw [->, line width=\redT{}] (u12) to [out=30, in=240] (u21);
\draw [->, line width=\redT{}] (u21) to [out=195, in=345] (u11);
\draw [->, line width=\redT{}] (xx) -- (u11);
\draw [->, line width=\redT{}] (u21) -- (xx);
}

\node at (3.5,-0.5) {$(b)$};
\end{tikzpicture} \hspace{0.85cm}
\begin{tikzpicture}[scale=0.45]
 \node (u11)  at (2,8) [vertexX] {$u_{1,1}$};
 \node (u12) at (0,6) [vertexX] {$u_{1,2}$};
 \node (h2) at (7.0,7.0) [vertexY] {$H_2$};
 \node (h3) at (1.0,1.0) [vertexY] {$H_3$};
 \node (h4) at (7.0,1.0) [vertexY] {$H_4$};

{\color{\blueC{}}
\draw [->, line width=\blueT{}] (u12) -- (h2);
\draw [->, line width=\blueT{}] (h2) to [out=240, in=30] (h3);
\draw [->, line width=\blueT{}] (h3) -- (h4);
\draw [->, line width=\blueT{}] (h4) -- (u11);
\draw [->, line width=\blueT{}] (u11) to [out=277, in=62] (h3);
\draw [->, line width=\blueT{}] (h3) -- (u12);
}

{\color{\redC{}}
\draw [->, line width=\redT{}] (u11) to [out=10, in=140] (h2);
\draw [->, line width=\redT{}] (h2) -- (h4);
\draw [->, line width=\redT{}] (h4) -- (u12);
\draw [->, line width=\redT{}] (u12) to [out=255, in=130] (h3);
\draw [->, line width=\redT{}] (h3) to [out=90, in=240] (u11);
}
\node at (3.3,-0.5) {$(c)$};
\end{tikzpicture}
\caption{\Good{} decompositions of different digraphs.  The \ColorXX{red}{thick} arcs form one strong spanning subdigraph and the \ColorXX{blue}{thin} arcs form the
other strong spanning subdigraph. \jbj{The arcs between $\{u_{1,1},u_{1,2}\}$ and the $H_i$'s \AY{in (c)} indicate the direction of all arcs between these sets.}} \label{fig_x1}
\end{figure}
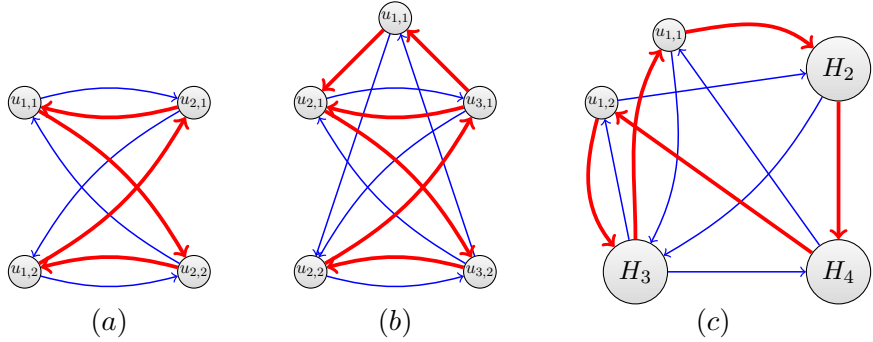

{\bf  Case 2.} $|V(T)| = 3$.

\2

If $|H_1|=|H_2|=|H_3|=2$ then we are done by Theorem~\ref{good-decompSemi}, so we may  without
loss of generality assume that $|H_1|=1$. 
As $T$ is a strong semicomplete digraph, it contains a Hamilton cycle, \jbj{by Camion's theorem (see \cite{Camion} or  \cite[Theorem 2.2.6]{bangTchapter}),} so we may assume that 
$u_1u_2,u_2u_3,u_3u_1 \in A(T)$. 
As $|H_1|=1$ and the vertices in $H_2$ have in-degree at least two we must then have
$u_3 u_2 \in A(T)$.

If $|H_2|=|H_3|=2$, then we note that $\induce{D}{H_2 \cup H_3}$ has a \good{} decomposition which can easily be extended to 
a \good{} decomposition of $D$ as can be seen in \AY{Figure~\ref{fig_x1}(b)}.
We may therefore  assume without loss of generality that 
$|H_1|=|H_2|=1$.  As the vertices in $H_3$ have in-degree at least two we must then have
$u_1 u_3 \in A(T)$.  

If $|H_3|=2$ then note that
 $u_{1,1} u_{3,1} u_{2,1} u_{3,2} u_{1,1}$ and
$u_{1,1} u_{3,2} u_{2,1} u_{3,1} u_{1,1}$ form arc-disjoint Hamilton cycles,
thereby proving that $D$ has a \good{} decomposition. 

The only remaining case is when 
$|H_1|=|H_2|=|H_3|=1$ and $u_2 u_1 \in A(T)$. 
However in this case $D$ has a \good{} decomposition as it consists of two $3$-cycles in the opposite directions.
This completes the proof of Case~2.

\2

{\bf  Case 3.} $T$ is $2$-arc-strong  and $|V(T)| \geq 4$.

\2

If  $T$ is not isomorphic to $S_4$, then we are done by Theorem~\ref{thm03} and Lemma \ref{lemo}.
Now assume that $T$ is isomorphic to $S_4$. If $|H_1|=|H_2|=|H_3|=|H_4|=1$, then $D$ is isomorphic to $S_4$ and $D \in {\cal D}_2$.
By symmetry we may therefore without loss of generality assume that $|H_1|=2$ \jbj{and } a \good{} decomposition of $D$ can be seen in 
\AY{Figure~\ref{fig_x1}(c)} (the arcs shown between all $u_{1,i}$'s
and $H_j$'s and between all $H_j$'s and $H_k$'s show the direction of all arcs between them), completing the proof of Case 3.

\2

{\bf  Case 4.} $T$ is not $2$-arc-strong and $|V(T)| \geq 4$.

\2
We will now prove by induction on $|V(D)|$ that $D$ has a \good{} decomposition. If $|V(D)|=4$ then $D$ is a semicomplete digraph and we are done by Theorem \ref{thm03}. Hence we may assume that $|V(D)|>4$
  and proceed to the induction step.\\

  Suppose first that $D$ has a vertex \AY{$z \in H_i$, where $|H_i|=2$,} so that $\hat{D}=D-z$ is 2-arc-strong. 
By induction $\hat{D}$ has a \good{} decomposition, unless it is one of the exceptions in the theorem.   \AY{
As we have assumed that $|V(H_i)|\leq 2$ for $i\in [t]$ we note that $\hat{D}$ in this case is either $S_4$ or $T^s_4[\overline{K}_{2},\overline{K}_{2},\overline{K}_{1},\overline{K}_{1}]$.
If $\hat{D}$ is isomorphic to $S_4$, then $T=\hat{D}$ and $T$ is $2$-arc-strong, a contradiction by the statement of Case 4.
Thus, we may assume that $\hat{D}$ is isomorphic to  $T^s_4[\overline{K}_{2},\overline{K}_{2},\overline{K}_{1},\overline{K}_{1}]$.
However in this case we can find a \good{} decomposition of $D = T^s_4[\overline{K}_{2},\overline{K}_{2},\overline{K}_{2},\overline{K}_{1}]$
as seen in Figure \ref{T_4[2,2,2,1]good}. Hence $\hat{D}$ has a \good{} decomposition and we are done by Lemma \ref{lemo}. 
}

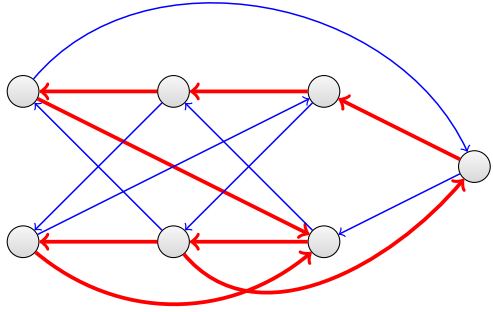
\begin{figure}
\begin{center}
\tikzstyle{vertexX}=[circle,draw, top color=gray!5, bottom color=gray!30, minimum size=20pt, scale=0.6, inner sep=0.7pt]
\tikzstyle{vertexY}=[circle,draw, minimum size=35pt, scale=1.1, inner sep=2.5pt]
\tikzstyle{vertexZ}=[rectangle, rounded corners,draw,minimum height=76pt, minimum width=30.5pt, scale=1.1, inner sep=2.5pt]

\begin{tikzpicture}[scale=0.5]
 \node (p3a) at (2.0,5.0) [vertexX] {};
 \node (p3b) at (2.0,1.0) [vertexX] {};
 \node (p2a) at (6.0,5.0) [vertexX] {};
 \node (p2b) at (6.0,1.0) [vertexX] {};
 \node (p1a) at (10.0,5.0) [vertexX] {};
 \node (p1b) at (10.0,1.0) [vertexX] {};
 \node (p0) at (14.0,3.0) [vertexX] {};

{\color{\redC{}}
\draw [->, line width=\redT{}] (p0) -- (p1a);
\draw [->, line width=\redT{}] (p1a) -- (p2a);
\draw [->, line width=\redT{}] (p2a) -- (p3a);

\draw [->, line width=\redT{}] (p3b) to [out=320, in=220] (p1b);
\draw [->, line width=\redT{}] (p3a) -- (p1b);

\draw [->, line width=\redT{}] (p1b) -- (p2b);
\draw [->, line width=\redT{}] (p2b) -- (p3b);

\draw [->, line width=\redT{}] (p2b) to [out=310, in=230] (p0);
}

{\color{\blueC{}}

\draw [->, line width=\blueT{}] (p0) -- (p1b);
\draw [->, line width=\blueT{}] (p1b) -- (p2a);
\draw [->, line width=\blueT{}] (p2a) -- (p3b);

\draw [->, line width=\blueT{}] (p3b) -- (p1a);

\draw [->, line width=\blueT{}] (p1a) -- (p2b);
\draw [->, line width=\blueT{}] (p2b) -- (p3a);

\draw [->, line width=\blueT{}] (p3a) to [out=50, in=115] (p0);

}
\end{tikzpicture}
\end{center}
\caption{A \good{} decomposition of $D = T^s_4[\overline{K}_{2},\overline{K}_{2},\overline{K}_{2},\overline{K}_{1}]$}\label{T_4[2,2,2,1]good}.
\end{figure}


Hence we may assume below that

\AY{
  \begin{equation}
    \label{not2AS}
    \begin{minipage}{7cm}
    \mbox{for every $H_i$ with $|H_i|=2$ and every $z\in V(H_i)$} \mbox{the digraph $D-z$ is not 2-arc-strong.}
    \end{minipage}
    \end{equation}
}

By Theorem~\ref{prop:nice}, $T$ has a nice vertex decomposition $(T_1, \dots , T_p).$ 
Let $(D_1,D_2,\ldots ,D_p)$ be the vertex decompostion of $D$ obtained by replacing 
each vertex $u_i$ of $T$ by the corresponding \AY{independent set}  $H_i$ (so $V(D_j)=\bigcup_{u_i\in V(T_j)}V(H_i)$).
Suppose that there is no \good{} decomposition of $D$. We now prove the following claims.

\2

{\bf  Claim \claimA{}.}  If $u_i u_j$ is a cut-arc in $T$ and $|H_j|=1$, then $V(T_1) = \{u_j\}.$

\2

{\bf  Proof of Claim \claimA{}:} Assume that $u_i u_j$ is a cut-arc in $T$ and $|H_j|=1$. 
Define $r$ such that $u_j \in T_r$.
Let $X = (V(D_1) \cup V(D_2) \cup \cdots \cup V(D_{r-1})) \setminus \{u_{j,1}\}$ and let
$Y = V(D_r) \cup V(D_{r+1}) \cup \cdots \cup V(D_p)$. Note that $Y \not= \emptyset$ as $V(D_r) \subseteq Y$.
Also note that there is no arc from $Y$ to $X$ in $D$ (as the only arcs out of $Y$ go to $H_j = \{u_{j,1}\}$).
Therefore if $X \not= \emptyset$ then we are done by Lemma~\ref{cutvertex}.
So, $X=\emptyset$, which implies that $V(T_1) = \{u_j\}$, completing the proof of Claim~\claimA{}.\qed

\2

{\bf  Claim \claimB{}.}  If $u_iu_j$ is a cut-arc in $T$ and $|H_i|=1$, then $T_p = \{u_i\}.$

\2

{\bf  Proof of Claim \claimB{}:} This can be proved analogously to Claim~\claimA{}. \qed

\2

{\bf  Claim \claimD{}.} If $|H_r|=2$, for some $r \in [t]$, then $u_r$ is incident with a cut-arc into $T_1$ or \AY{a cut-arc} out of $T_p$ (or both).

This implies that for every $u_q\in V(T)$ that is not incident to a cut-arc we have $|V(H_q)|=1$.
  
\2

{\bf  Proof of Claim \claimD{}:} 
\AY{
Let $|H_r|=2$, for some $r \in [t]$. By Lemma~\ref{endofcutarc} we note that there exists a $q\neq r$ such that one of the following holds.

\begin{description}
\item[(i)] $u_ru_q$ is a cut-arc of $T$ and $N^-(V(H_q))=V(H_r)$, or
\item[(ii)] $u_qu_r$ is a cut-arc of $T$ and $N^+(V(H_q))=V(H_r)$.
\end{description}

Assume without loss of generality that (i) above holds. Therefore $u_r$ is incident with the cut-arc
$u_ru_q$ in $T$. As $N^-(V(H_q))=V(H_r)$, we note that $d_T^-(u_q)=1$. If $u_q \in T_1$, then we are done, so assume that this is not the
case. However, as $u_ru_q$ is a cut-arc in $T$ observe that $u_r \not\in T_1$, which implies that $u_q$ dominates $T_1$ (as (i) holds).
This implies that $|T_1|=1$, and assuming that $T_1=\{u_1\}$, we note that $u_q u_1$ is the cut-arc into $T_1$ in $T$. 
Therefore $N_T^-(u_1) = \{u_q\}$.
As $u_ru_q$ is a cut-arc in $T$ we note that $u_q \not\in T_p$, which by Claims~\claimA{} and \claimB{} implies that $|H_q|=2$ (as 
$u_qu_1$ is a cut-arc in $T$). 

We will now show that $|H_1|=1$. For the sake of contradiction assume that $|H_1|=2$, which by Lemma~\ref{endofcutarc}(ii), implies
that $N^+(H_q)=H_1$. However this implies that $d_T^+(u_q)=1$ and from above $d_T^-(u_q)=1$, implying that $|V(T)|=3$, contradicting the 
fact that $|V(T)| \geq 4$. Therefore $|H_1|=1$. As $T_1=\{u_1\}$ we note that $d_T^-(u_1)=1$.

To summarize, we have now shown the following.

\begin{itemize}
 \item[(I)] $H_1 = \{u_{1,1}\},$ $H_q = \{u_{q,1},u_{q,2}\}$ and $H_r = \{u_{r,1},u_{r,2}\}$ in $D$.
 \item[(II)] $u_r u_q$ and $u_q u_1$ are cut-arcs in $T$.
 \item[(III)] $N_T^-(u_1) = \{u_q\}$ and $N_T^-(u_q) = \{u_r\}$.
\end{itemize}

We now consider $D' = D - u_{1,1}$.  By Lemma~\ref{cutvertex} we note that $D$ contains no cut-vertices 
and therefore $D'$ is strongly connected.  If $D'$ has a \good{} decomposition then this can easily be 
extended to a \good{} decomposition of $D$, as $u_{1,1}$ has at least two arcs into $D'$ and at least two arcs out of $D'$.
We may therefore assume that $D'$ has no \good{} decomposition.

If $D'$ is $2$-arc-connected, this implies that $D'$ is one of our exceptions. In this case $D'$ is either $\overrightarrow{C}_3[\overline{K}_2,\overline{K}_2,\overline{K}_2]$ or $\overrightarrow{C}_3[\overline{K}_2,\overline{K}_2,\overrightarrow{P}_2]$, since $S_4$ is semicomplete. If $D'=\overrightarrow{C}_3[\overline{K}_2,\overline{K}_2,\overline{K}_2]$, then it follows from (III) that $D=T^s_4[\overline{K}_1,\overline{K}_2,\overline{K}_2,\overline{K}_2]$ and we obtain a good decomposition of $D$ using the decomposition in Figure \ref{T_4[2,2,2,1]good} and then reversing all arcs. Suppose now that
$D'=\overrightarrow{C}_3[\overline{K}_2,\overline{K}_2,\overrightarrow{P}_2]$. Then $D$ contains  $T^s_4[\overline{K}_1,\overline{K}_2,\overline{K}_2,\overline{K}_2]$ as a spanning subdigraph and hence it has a \good{} decomposition.\\

So we may now assume that $D'$ is not $2$-arc-connected and we will let $(S,\overline{S})$ be a partition of $V(D')$ with 
exactly one arc from $S$ to $\overline{S}$. We will now show that $d_{D'}^-(u_{r,1})=1$ (and $d_{D'}^-(u_{r,2})=1$).
In order to do this we consider the three possible placements of the vertices $u_{q,1}$ and $u_{q,2}$ in the partition
$(S,\overline{S})$.

\2

{\em Case \claimD{}.1: $\{u_{q,1},u_{q,2}\} \subseteq S$.} If $\overline{S}$ contains any vertex not in $V(H_r)$  then 
there are at least two arcs from $S$ to $\overline{S}$ (comming from  $u_{q,1}$ and $u_{q,2}$), a contradiction. 
Therefore, $\overline{S} \subseteq H_r$. Without loss of generality $u_{r,1} \in \overline{S}$ and $x u_{r,1}$ is the
arc from $S$ to $\overline{S}$. Considering $u_{r,1}$ we note that in $D'$ it only has one arc into it (from $x$), 
implying that $d_{D'}^-(u_{r,1})=1$ as desired. 

\2

{\em Case \claimD{}.2: $\{u_{q,1},u_{q,2}\} \subseteq \overline{S}$.} 
Adding $u_{1,1}$ to $\overline{S}$ we note that there is still only one arc from $S$ to $\overline{S}$ (as $u_{1,1}$ 
only has arcs into it from $u_{q,1}$ and $u_{q,2}$), a contradiction to $D$ being $2$-arc-strong.

\2

{\em Case \claimD{}.3: $|\{u_{q,1},u_{q,2}\} \cap S|=1.$} Without loss of generality assume that 
$u_{q,1} \in S$ and $u_{q,2} \in \overline{S}$. As there is only one arc from $S$ to $\overline{S}$ we note that
$u_{r,1}$ or $u_{r,2}$ must belong to $\overline{S}$. Without loss of generality assume that $u_{r,2} \in \overline{S}$.
For the sake of contradiction assume that $d_{D'}^-(u_{r,1}) \geq 2$ (and therefore $d_{D'}^-(u_{r,2}) \geq 2$) and
let $z_1,z_2 \in N_{D'}^-(u_{r,1})$ (and therefore $z_1,z_2 \in N_{D'}^-(u_{r,2})$) be arbitrary.

If $z_i \in S$, then we note that $z_i u_{r,2}$ is an arc from $S$ to $\overline{S}$ and 
if $z_i \in \overline{S}$ then either $z_i=u_{q,2}$ or $u_{q,1} z_i$ is an arc from $S$ to $\overline{S}$ for $i=1,2$.
As there is only one arc from $S$ to $\overline{S}$ we note that $z_1=u_{q,2}$ or $z_2=u_{q,2}$. Without loss of generality
we may assume that $z_1=u_{q,1}$ and $z_2=u_{q,2}$ (as if $u_{q,2}$ dominates $u_{r,1}$ then so does $u_{q,1}$).
However if $u_{r,1} \in S$ then both $u_{r,1} u_{q,2}$ and $u_{q,1} u_{r,2}$ go from $S$ to $\overline{S}$ and
if $u_{r,1} \in \overline{S}$ then both $u_{q,1} u_{r,1}$ and $u_{q,1} u_{r,2}$ go from $S$ to $\overline{S}$, a 
contradiction.  This completes Case \claimD{}.3.

\2

We have now shown that $d_{D'}^-(u_{r,1})=1$, so we may define $z$, such that $N_{D'}^-(u_{r,1})=\{u_{z,1}\}$.
Note that $|H_z|=1$.  Let $Y = H_1 \cup H_q \cup H_r \cup H_z$. We will show that $V(D)=Y$, so assume for the sake
of contradiction that there exists a vertex $y \in V(D) \setminus Y$. 
Then there is no path from $y$ to $Y \setminus \{u_{z,1}\}$ in $D - u_{z,1}$, as 
$N_T^-(u_1) = \{u_q\}$ and $N_T^-(u_q) = \{u_r\}$ and $N_T^-(u_r)=\{u_z\}$ (so all arcs into $Y  \setminus \{u_{z,1}\}$
come from $u_{z,1}$). 
This implies that $u_{z,1}$ is a cut-vertex in $D$, a contradiction by Lemma~\ref{cutvertex}.
Therefore we must have $V(D)=Y$.

As $|V(T)| \geq 4$, we note that $H_z$, $H_1$, $H_q$ and $H_r$ are distinct.
However in this case $D$ is the exception
$\overrightarrow{C}_3[\overline{K}_2,\overline{K}_2,\overrightarrow{P}_2]$.
This completes the proof of Claim~\claimD{}. \qed
 }

\2

{\bf  Claim \claimC{}.} \AY{$T$ has at most three cut-arcs.

\2

{\bf  Proof of Claim \claimC{}:} Let $u_iu_j$ be a cut-arc in $T$ with $1 < j < i < p$. 
By Claims~\claimA{} and \claimB{} we note that $|H_j|=|H_i|=2$. By Claim~\claimD{} we note that 
$u_j$ is incident with a cut-arc into $T_1$ and $u_i$ is incident with a cut-arc from $T_p$.
This implies that there are only these three cut-arcs in this case.  Furthermore, if there is no 
cut-arc, $u_iu_j$, in $T$ with $1 < j < i < p$ then there are at most two cut-arcs in $T$ (one into $T_1$ 
and one out of $T_p$). \qed
}

\2

The remaining part of the proof is split into three cases, covering the number of possible cut-arcs in $T$ according to Claim \claimC{}.\\

{\bf Case 4.1}. $T$ has exactly one cut-arc $u_pu_1$.\\

\AY{
Let $T'$ be the semicomplete multigraph that we obtain by adding an extra copy of the arc $u_pu_1$ to $T$ (so $T'$ has exactly one pair of parallel arcs). As $u_pu_1$ was the only cut-arc in $T$, we note that $T'$ is 2-arc strong.
By the statement of Case 4 we note that $T$ is not $2$-arc-strong, and therefore not isomorphic to $S_4$, 
implying that $T'$ is not one of the exceptions in 
Theorem~\ref{sdmulti}. Therefore $T'$ contains a \good{} decomposition  $(R_1,R_2)$. 

First consider the case when $|H_p|=|H_1|=2$.  
Let $R_1',R_2'$ be the arc-disjoint spanning subdigraphs of $D$ that we 
obtain by replacing the vertex $u_p$ by $\{u_{p,1},u_{p,2}\}$ and the vertex $u_1$
by $\{u_{1,1},u_{1,2}\}$. That is, if $xu_p$ ($u_py$) is an arc of $R_i$, then 
$R_i'$ contains the arcs $xu_{p,1},xu_{p,2}$ ($u_{p,1}y,u_{p,2}y$) and analogously for arcs entering and leaving $u_1$.
This is well-defined for all arcs apart from the ones from $\{u_{p,1},u_{p,2}\}$ to $\{u_{1,1},u_{1,2}\}$, for these we let
$u_{p,1} u_{1,1}$ and $u_{p,2} u_{1,2}$ belong to $R_1'$ and 
$u_{p,1} u_{1,2}$ and $u_{p,2} u_{1,1}$ belong to $R_2'$.
We will now show that $(R_1',R_2')$ is a \good{} decomposition of $D$.

As there is a path from $u_1$ to $u_p$ in $R_i$ ($ i\in [2]$), we note that for $j\in [2]$ the vertex $u_{p,j}$
 can reach every vertex in $\{u_{1,1},u_{1,2}\}$
in $R_i'$, either by a direct arc, or by an arc from $H_p$ to $H_1$ 
followed by the equivalent of a $(u_1,u_p)$-path in $R_i$ followed by another arc from $H_p$ to $H_1$.
(For example, in $R'_1$ the vertex $u_{p,1}$ can reach $u_{1,2}$ via the arc $u_{p,1}u_{1,2}$ followed by a $(u_{1,1},u_{p,2})$-path in $R_1$ and finaly the arc $u_{p,2}u_{1,2}$).
Therefore we have all the same connections
in $R_1'$ and $R_2'$ as in $R_1$ and $R_2$, completing the proof of the case when $|H_p|=|H_1|=2$. 

We may therefore without loss of generality assume that $|H_1|=1$.  As $D$ is $2$-arc-strong and $u_{1,1}$ is not a cut-vertex we note that $|H_p|=2$, $V(D_1)=\{u_{1,1}\}$ and there exists a vertex $u_{y,1} \in N_D^+(u_{p,1}) \setminus \{u_{1,1}\}$. Without loss of
generality assume that $u_p u_y \in A(R_1)$. As $D - u_{p,2} = T$ we can assign every arc of $D - u_{p,2}$ to $R_i'$ if
and only if it was assigned to $R_i$ in $T''$,
except the arc from $u_{p,1}$ to $u_{1,1}$ which gets assigned
to $R_2'$. Let $u_{x} u_p$ be the last arc on a path from $u_y$ to $u_p$ in $R_1$. Now let the arcs $u_{x,1} u_{p,2}$ 
and $u_{p,2} u_{1,1}$ belong to $R_1'$. Now we note that $R_1'$ is a strong spanning subdigraph of $D$, as there exists 
a path from both $u_{p,1}$ and $u_{p,2}$ to $u_{1,1}$ (and therefore all paths in $R_1$ also work for $R_1'$). 
Adding any arc into $u_{p,2}$, different from $u_{x,1} u_{p,2}$ and any arc out of $u_{p,2}$, different from 
$u_{p,2} u_{1,1}$ to $R_2'$ makes $R_2'$ into a strong spanning subdigraph of $D'$ (as it was already
strong in $D-u_{p,2}$).  Therefore $(R_1',R_2')$ is a \good{} decomposition of $D$. 
}

\2

{\bf Case 4.2}. $T$ has exactly two cut-arcs $u_pu_h$ and $u_ku_1$.
\AY{By Claim~\claimD{} we note that $|H_1|=|H_p|=1$ and as $D$ is $2$-arc-strong we note that $|H_h|=|H_k|=2$.
By Claim~\claimA{} and \claimB{} we note that $|T_1|=|T_p|=1$. This and the fact that $D$ has no cut-vertex implies that $|V(D_1)|=|V(D_p)|=1$.}
There are 3 subcases to consider: $u_k=u_h$, $u_k,u_h$ are distinct but  $u_k,u_h\in V(T_i)$ for some $i\in [p]$ and finally
the case where $u_k\in V(T_j),u_h\in V(T_i)$ where $i<j$.\\

{\bf Case 4.2.1}. $u_k=u_h$. Let the index $i$ be chosen so that $u_k\in V(T_i)$. 
\AY{As $D$ has no cut-vertex we note that $i=2$ and $p=3$. As
$|T_1|=|T_3|=1$ and $|V(T)| \geq 4$ we must have $|T_2| \geq 2$. }
Therefore the set $W=V(D_2)-\{u_{k,1},u_{k,2}\}$ contains at least one vertex. 
As $V(T_2)$ is strong the digraph $D'_2=D_2-\{u_{k,1}\}$ is strong.
  If $V(T_2)=\{u_k,u_r\}$ for some $r$ (that is, $V(T)=\{u_1,u_k,u_r,u_p\}$) then the hamiltonian cycle
  $u_{1,1}u_{p,1}u_{k,1}u_{r,1}u_{k,2}u_{1,1}$ (see (a) below) is arc-disjoint from the strong spanning subdigraph whose arc set is the arcs of the two paths $u_{p,1}u_{k,2}u_{r,1}u_{k,1}u_{1,1}$ and $u_{1,1}u_{r,1}u_{p,1}$ (see (b) below), showing that $D$ has a \good{} decomposition.

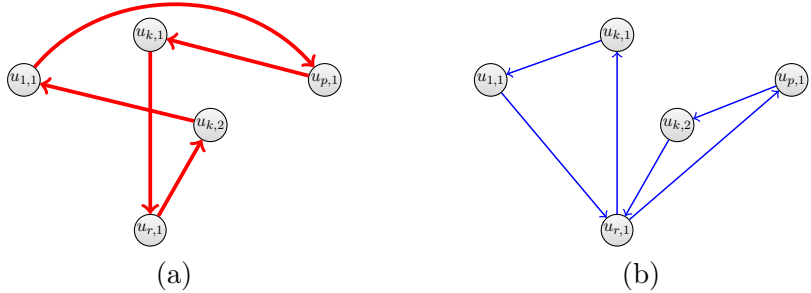
\begin{center}
\tikzstyle{vertexX}=[circle,draw, top color=gray!5, bottom color=gray!30, minimum size=16pt, scale=0.6, inner sep=0.5pt]
\tikzstyle{vertexY}=[circle,draw, top color=gray!5, bottom color=gray!30, minimum size=20pt, scale=0.7, inner sep=1.5pt]
\begin{tikzpicture}[scale=0.4]
 \node (u1) at (1.0,7.0) [vertexX] {$u_{1,1}$};
 \node (k1) at (5.2,8.5) [vertexX] {$u_{k,1}$};
 \node (k2) at (7.2,5.5) [vertexX] {$u_{k,2}$};
 \node (p1) at (11.0,7.0) [vertexX] {$u_{p,1}$};
 \node (r1) at (5.2,2.0) [vertexX] {$u_{r,1}$};

{\color{\redC{}}
\draw [->, line width=\redT{}] (u1) to [out=50, in=130] (p1);
\draw [->, line width=\redT{}] (p1) -- (k1);
\draw [->, line width=\redT{}] (k1) -- (r1);
\draw [->, line width=\redT{}] (r1) -- (k2);
\draw [->, line width=\redT{}] (k2) -- (u1);
}

\node at (6,0.5) {(a)};
\end{tikzpicture} \hspace{1.5cm}
\begin{tikzpicture}[scale=0.4]
 \node (u1) at (1.0,7.0) [vertexX] {$u_{1,1}$};
 \node (k1) at (5.2,8.5) [vertexX] {$u_{k,1}$};
 \node (k2) at (7.2,5.5) [vertexX] {$u_{k,2}$};
 \node (p1) at (11.0,7.0) [vertexX] {$u_{p,1}$};
 \node (r1) at (5.2,2.0) [vertexX] {$u_{r,1}$};


{\color{\blueC{}}
\draw [->, line width=\blueT{}] (p1) -- (k2);
\draw [->, line width=\blueT{}] (k2) -- (r1);
\draw [->, line width=\blueT{}] (r1) -- (k1);
\draw [->, line width=\blueT{}] (k1) -- (u1);
\draw [->, line width=\blueT{}] (u1) -- (r1);
\draw [->, line width=\blueT{}] (r1) -- (p1);
}
\node at (6,0.5) {(b)};
\end{tikzpicture}

\end{center}

  Suppose now that $|V(T_2)|>2$. Let $v$  be an in-neighbour of $u_{k,1}$ in $V(D_2)$ and let $w\neq v$ be an out-neighbour of $u_{k,1}$ in $V(D_2)$. Let $D'_2=D_2-u_{k,1}$ and note that $D'_2$ is strong.
  Now let the two spanning digraphs $G_1=(V,A_1),G_2=(V,A_2)$ contain the following arcs (see Figure \ref{figx1a}).
  \begin{itemize}
    
  \item $A_1=\{u_{p,1}u_{k,1},u_{k,1}u_{1,1},u_{1,1}w,vu_{p,1}\}\cup{}A(D'_2)$
  \item $A_2=\{u_{p,1}u_{k,2},u_{k,2}u_{1,1},u_{1,1}v,vu_{k,1},u_{k,1}w,wu_{p,1}\}\cup\{u_{1,1}z,zu_{p,1}|z\in V(D'_2)-\{u_{k,2},v,w\}$
    \end{itemize}

    It is easy to verify that $G_1,G_2$ are arc-disjoint strong spanning subdigraphs of $D$.

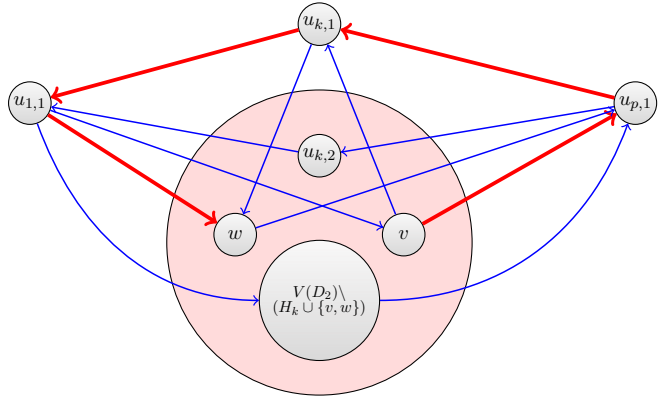
\begin{figure}[h!]
\begin{center}
  \tikzstyle{vertexX}=[circle,draw, top color=gray!5, bottom color=gray!30, minimum size=23pt, scale=0.7, inner sep=0.5pt]
\tikzstyle{vertexY}=[circle,draw, top color=gray!5, bottom color=gray!30, minimum size=65pt, scale=0.7, inner sep=1.5pt]
\tikzstyle{vertexZ}=[circle,draw, top color=red!14, bottom color=red!14, minimum size=165pt, scale=0.7, inner sep=1.5pt]
\tikzstyle{vertexW}=[circle,draw, top color=gray!10, bottom color=gray!10, minimum size=165pt, scale=0.7, inner sep=1.5pt]
\begin{tikzpicture}[scale=0.35]
\ColorXX{\node (DD) at (12.0,4.7) [vertexZ] {};}{\node (DD) at (12.0,4.7) [vertexW] {};}

 \node (u11) at (1,10) [vertexX] {$u_{1,1}$};
 \node (uk1) at (12,13) [vertexX] {$u_{k,1}$};
 \node (uk2) at (12,8) [vertexX] {$u_{k,2}$};
 \node (up1) at (24,10) [vertexX] {$u_{p,1}$};

 \node (w) at (8.8,5.0) [vertexX] {$w$};
 \node (v) at (15.2,5.0) [vertexX] {$v$};
 \node (rest) at (12.0,2.5) [vertexY] {};

{\color{\redC{}}
\draw [->, line width=\redT{}] (up1) -- (uk1);
\draw [->, line width=\redT{}] (uk1) -- (u11);
\draw [->, line width=\redT{}] (u11) -- (w);
\draw [->, line width=\redT{}] (v) -- (up1);
}

{\color{\blueC{}}
 \draw [->, line width=\blueT{}] (up1) -- (uk2);
 \draw [->, line width=\blueT{}] (uk2) -- (u11);

 \draw [->, line width=\blueT{}] (v) -- (uk1);
 \draw [->, line width=\blueT{}] (uk1) -- (w);
 \draw [->, line width=\blueT{}] (u11) -- (v);
 \draw [->, line width=\blueT{}] (w) -- (up1);

 \draw [->, line width=\blueT{}] (u11) to [out=290, in=180] (rest);
 \draw [->, line width=\blueT{}] (rest) to [out=0, in=250] (up1);
 }
 \node at (12.0,2.8) [scale=0.55] {$V(D_2)\setminus$};
 \node at (12.0,2.2) [scale=0.55] {$(H_k \cup \{v,w\})$};

\end{tikzpicture} 

  \caption{The \good{} decomposition used in Case 4.2.1, where all arcs within the big \ColorXX{red}{gray} circle are \ColorXX{red}{thick} arcs.}\label{figx1a}
\end{center}
\end{figure}

  \2

  {\bf Case 4.2.2} $u_k$ and $u_h$ are distinct but belong to the same $T_i$.\\

\AY{As in the proof of Case 4.2.1, we note that $i=2$ and $p=3$  (as $D$ has no cut-vertex) and $|V(D_1)|=|V(D_3)|=1$.
}

 As $u_pu_h$ and $u_ku_1$ are the only cut-arcs of $T$ it follows from Menger's theorem that there are two arc-disjoint $(u_h,u_k)$-paths $P_1,P_2$ in 
\AY{$T_2$}. 
For $i\in [2]$ let $A'_i$ be the arcs of \AY{$D_2$} that correspond to $A(P_i)$, that is, 
we replace the first  arc $u_hv$ (last arc $v'u_k$) of $P_i$ by the two arcs $u_{h,1}v,u_{h,2}v$ (respectively, $v'u_{k,1},v'u_{k,2}$). Recall that, by Claim \claimD{}, we have \AY{$|H_g|=1$} when $u_g$ is not incident to a cut-arc of $T$ so $A(P_i)$ corresponds exactly to $A'_i$ in $D_2$. 

\AY{We will now construct $F_1$ and $F_2$ as follows. Let $X = V(D_2) \setminus (H_k \cup H_h)$. Initially let $F_1$ and $F_2$ consist of the following arcs (see Figure \ref{figx2A}):

\begin{itemize}
 \item $A(F_1)$ initially consists of the arcs $\{u_{p,1}u_{h,1},u_{k,2}u_{1,1},u_{1,1}u_{h,2},u_{k,1}u_{p,1}\}$ and all arcs of $A'_1$. 
 \item $A(F_2)$ initially consists of the arcs $\{u_{p,1}u_{h,2},u_{k,2}u_{p,1},u_{1,1}u_{h,1},u_{k,1}u_{1,1}\}$ and all arcs of $A'_2$. 
\end{itemize}
 
}

\begin{figure}[h!]
  \begin{center}
    \tikzstyle{vertexX}=[circle,draw, top color=gray!5, bottom color=gray!30, minimum size=17pt, scale=0.7, inner sep=0.5pt]
\tikzstyle{vertexY}=[circle,draw, top color=gray!5, bottom color=gray!30, minimum size=65pt, scale=0.7, inner sep=1.5pt]
\tikzstyle{vertexZ}=[rectangle, rounded corners,draw,minimum height=80pt, minimum width=80pt, scale=1.1, inner sep=2.5pt]

\begin{tikzpicture}[scale=0.6]
 \node (DD) at (9,7) [vertexZ] {$X$};

 \node (u11) at (1,7) [vertexX] {$u_{1,1}$};

 \node (uh1) at (11,11) [vertexX] {$u_{h,1}$};
 \node (uh2) at (7,11) [vertexX] {$u_{h,2}$};

 \node (uk1) at (11,3) [vertexX] {$u_{k,1}$};
 \node (uk2) at (7,3) [vertexX] {$u_{k,2}$};

 \node (up1) at (17,7) [vertexX] {$u_{p,1}$};

 \node (kr) at (10.5,5.5) [vertexX] {};
 \node (kb) at (7.5,5.5) [vertexX] {};
 \node (hr) at (10.5,8.5) [vertexX] {};
 \node (hb) at (7.5,8.5) [vertexX] {};

{\color{\redC{}}
\draw [->, line width=\redT{}] (uh1) -- (hr);
\draw [->, line width=\redT{}] (uh2) -- (hr);
\draw [->, line width=\redT{}] (kr) -- (uk1);
\draw [->, line width=\redT{}] (kr) -- (uk2);
\draw [->, line width=\redT{}] (up1) -- (uh1);
\draw [->, line width=\redT{}] (u11) -- (uh2);
\draw [->, line width=\redT{}] (uk1) -- (up1);
\draw [->, line width=\redT{}] (uk2) -- (u11);
\draw[->, snake=coil,segment aspect=0, line width=\redT{}] (hr)  -- (kr);
\node at (11.0,7) [scale=0.75] {$P_1$};

}

{\color{\blueC{}}
\draw [->, line width=\blueT{}] (uh1) -- (hb);
\draw [->, line width=\blueT{}] (uh2) -- (hb);
\draw [->, line width=\blueT{}] (kb) -- (uk1);
\draw [->, line width=\blueT{}] (kb) -- (uk2);
\draw [->, line width=\blueT{}] (u11) to [out=70, in=135] (uh1);
\draw [->, line width=\blueT{}] (up1) to [out=110, in=45] (uh2);

\draw [->, line width=\blueT{}] (uk1) to [out=225, in=290] (u11);
\draw [->, line width=\blueT{}] (uk2) to [out=315, in=250] (up1);
\draw[->, snake=coil,segment aspect=0, line width=\blueT{}] (hb)  -- (kb);
\node at (7.0,7) [scale=0.75] {$P_2$};
 }

\end{tikzpicture} 

  \caption{The initial assignments of arcs to $F_1$ and $F_2$ in Case 4.2.2.}\label{figx2A}
\end{center}
\end{figure}
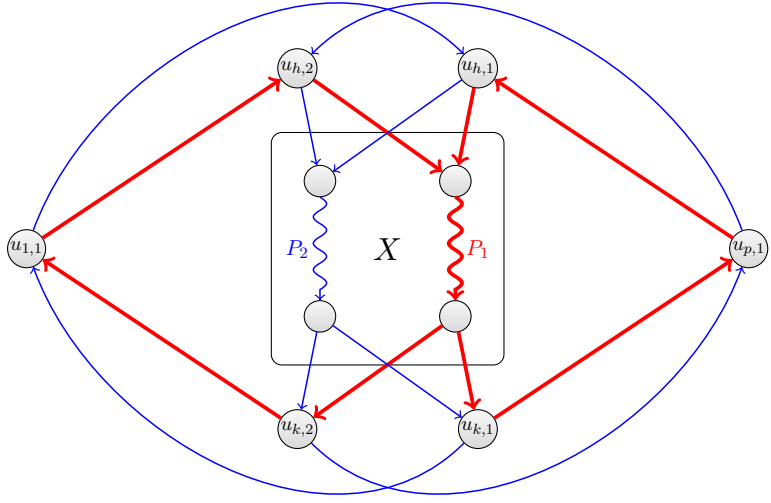

\AY{Now for every vertex $x \in X$ we add the following arcs to $F_1$ and $F_2$.

\begin{itemize}
 \item If $x \not\in V(P_1)$ then add the arcs $u_{1,1}x$ and $xu_{p,1}$ to $A(F_1)$.
 \item If $x \in V(P_1)$ then add the arcs $u_{1,1}x$ and $xu_{p,1}$ to $A(F_2)$.
\end{itemize}

Finally we add all arcs not assigned to any $F_i$ yet to $A(F_2)$. It is easy to check that $F_1$ is a 
strong spanning subdigraph of $D$.  In order to show that $F_2$ is also a strong spanning subdigraph of $D$ 
we consider any $x \in X$ and will show that $x$ has a path to and from $\{u_{p,1},u_{1,1}\}$ in $F_2$.
If $x \in V(P_1) \cup V(P_2)$, then this is clearly the case by the construction above (as either the arcs $u_{1,1}x$ and $xu_{p,1}$
belong to $F_2$ or $x \in V(P_2)$). So assume 
that $x \not\in V(P_1) \cup V(P_2)$. In this case any path from $x$ to $V(P_1) \cup V(P_2)$ in $D_2$, and
any path to $x$ from $V(P_1) \cup V(P_2)$, belongs
to $F_2$, so we are done as all vertices in $V(P_1) \cup V(P_2)$ have a path to and from $\{u_{p,1},u_{1,1}\}$ in $F_2$.
Therefore $(F_1,F_2)$ is a \good{} decomposition of $D$.
This completes the proof of Case 4.2.2.
}

  \2

  {\bf Case 4.2.3} There are indices $1<i<j<p$ such that $u_h\in V(T_i)$ and $u_k\in V(T_j)$.\\

\AY{ 
Recall that $|V(H_1)|=|V(H_p)|=1$ and $|T_1|=|T_p|=1$, implying that we must have $V(D_1)=\{u_{1,1}\}$ and $V(D_p)=\{u_{p,1}\}$. As $D$ has no cut vertex we must furthermore have $i=2$ and $j=p-1$. }
If $V(D)=\{u_{1,1},u_{h,1},u_{h,2},u_{k,1},u_{k,2},u_{p,1}\}$, then $T$ has another cut-arc, namely $u_2u_3$, 
contradicting that we are in Case 4.2. 
Thus we can choose a vertex $z\in V(D)-\{u_{1,1},u_{h,1},u_{h,2},u_{k,1},u_{k,2},u_{p,1}\}$ so that $z$ is an out-neighbour of $u_{h,1},u_{h,2}$ and an in-neighbour of $u_{k,1},u_{k,2}$. Let  $U=\{z,u_{1,1},u_{h,1},u_{h,2},u_{k,1},u_{k,2},u_{p,1}\}$ and note that every vertex of $V(D)-U$ has at least two in-neighbours and at least two out-neighbours in $U$ so it suffices to give a \good{} decomposition for 
$D[U]$. Such a decomposition $H'_1,H'_2$ is shown in Figure \ref{figx4}. 
The two arc-disjoint digraphs contain  the following arcs: $A(H'_1)$ contains the arcs of the 6-cycle $u_{p,1}u_{h,1}u_{k,1}u_{1,1}u_{h,2}u_{k,2}u_{p,1}$ and the two arcs $u_{1,1}z,zu_{p,1}$ and $A(H'_2)$ contains the arcs of the 5-cycle $u_{k,2}u_{1,1}u_{p,1}u_{h,2}zu_{k,2}$ and the arcs of the paths $u_{1,1}u_{h,1}u_{k,2}$ and $u_{h,2}u_{k,1}u_{p,1}$.

\begin{figure}[h!]
\begin{center}
  \tikzstyle{vertexX}=[circle,draw, top color=gray!5, bottom color=gray!30, minimum size=23pt, scale=0.7, inner sep=0.5pt]
\tikzstyle{vertexY}=[circle,draw, top color=gray!5, bottom color=gray!30, minimum size=20pt, scale=0.7, inner sep=1.5pt]
\begin{tikzpicture}[scale=0.35]
 \node (u11) at (1.0,8.0) [vertexX] {$u_{1,1}$};
 \node (uh1) at (8.0,11.0) [vertexX] {$u_{h,1}$};
 \node (uh2) at (8.0,5.0) [vertexX] {$u_{h,2}$};
 \node (uk1) at (16.0,11.0) [vertexX] {$u_{k,1}$};
 \node (uk2) at (16.0,5.0) [vertexX] {$u_{k,2}$};
 \node (up1) at (23.0,8.0) [vertexX] {$u_{p_1}$};
 \node (z) at (12.0,1.0) [vertexX] {$z$};

{\color{\redC{}}
\draw [->, line width=\redT{}] (up1) to [out=110, in=45] (uh1);
\draw [->, line width=\redT{}] (uk1) to [out=125, in=70] (u11);
\draw [->, line width=\redT{}] (z) to [out=0, in=230] (up1);
\draw [->, line width=\redT{}] (u11) to [out=310, in=180] (z);
\draw [->, line width=\redT{}] (u11) -- (uh2);
\draw [->, line width=\redT{}] (uh2) -- (uk2);
\draw [->, line width=\redT{}] (uk2) -- (up1);
\draw [->, line width=\redT{}] (uh1) -- (uk1);
}

{\color{\blueC{}}
 \draw [->, line width=\blueT{}] (u11) -- (uh1);
 \draw [->, line width=\blueT{}] (uh1) -- (uk2);

 \draw [->, line width=\blueT{}] (uk2) to [out=160, in=0] (u11);

 \draw [->, line width=\blueT{}] (up1) to [out=180, in=20] (uh2);

 \draw [->, line width=\blueT{}] (uh2) -- (uk1);
 \draw [->, line width=\blueT{}] (uk1) -- (up1);
 \draw [->, line width=\blueT{}] (uh2) -- (z);
 \draw [->, line width=\blueT{}] (z) -- (uk2);
 \draw [line width=\blueT{}] (u11) to [out=90, in=180] (5,15);
 \draw [line width=\blueT{}] (5,15) to [out=0, in=180] (19,15);
 \draw [->, line width=\blueT{}] (19,15) to [out=0, in=90] (up1);
 }
\node at (12,0.5) {\mbox{ }};
\end{tikzpicture} 

  \caption{The \good{} decomposition used in Case 4.2.3.}\label{figx4}
\end{center}
\end{figure}
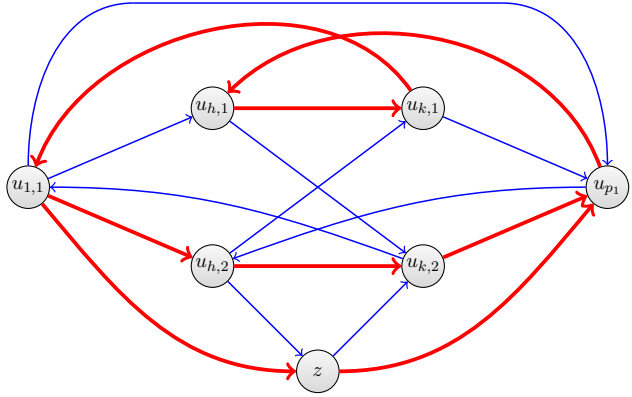

\2
  
{\bf Case 4.3}. $T$ has three cut-arcs $u_pu_h,u_hu_k,u_ku_1$.\\

\AY{Recall that $|T_1|=|T_p|=1$ and $|H_1|=|H_p|=1$, which implies that $V(D_1)=\{u_{1,1}\}$ and $V(D_p)=\{u_{p,1}\}$. }
If $V(D)=\{u_{1,1},u_{k,1},u_{k,2},u_{h,1},u_{h,2},u_{p,1}\}$, then $D$ is isomorphic to
$T_4^s[\overline{K}_2,\overline{K}_2,\overline{K}_1,\overline{K}_1]=\overrightarrow{C}_3[\overline{K}_2,\overline{K}_2,\overrightarrow{P}_2]$. 
Thus, we may assume that $|V(D)|\geq 7.$ Now we can choose a vertex $w$ which is an out-neighbour of $u_{k,1},u_{k,2}$ and an in-neighbour of $u_{h,1},u_{h,2}$. As above it suffices to show that the subdigraph induced by $\{w,u_{1,1},u_{k,1},u_{k,2},u_{h,1},u_{h,2},u_{p,1}\}$ has a \good{} decomposition. This follows from the fact that the subdigraphs
$\bar{H}_1,\bar{H}_2$ are strong and arc-disjoint where $A(\bar{H}_1)$ contains the arcs of the 5-cycle
$u_{1,1}wu_{p,1}u_{h,2}u_{k,2}u_{1,1}$ and the path $u_{1,1}u_{h,1}u_{k,1}u_{p,1}$ and $A(\bar{H}_2)$ is the 7-cycle $u_{1,1}u_{p,1}u_{h,1}u_{k,2}wu_{h,2}u_{k,1}u_{1,1}$ (see Figure~\ref{figx8}).

\begin{figure}[h!]
\begin{center}
  \tikzstyle{vertexX}=[circle,draw, top color=gray!5, bottom color=gray!30, minimum size=23pt, scale=0.7, inner sep=0.5pt]
\tikzstyle{vertexY}=[circle,draw, top color=gray!5, bottom color=gray!30, minimum size=20pt, scale=0.7, inner sep=1.5pt]
\begin{tikzpicture}[scale=0.35]
 \node (u11) at (1.0,8.0) [vertexX] {$u_{1,1}$};
 \node (uk1) at (8.0,11.0) [vertexX] {$u_{k,1}$};
 \node (uk2) at (8.0,5.0) [vertexX] {$u_{k,2}$};
 \node (uh1) at (16.0,11.0) [vertexX] {$u_{h,1}$};
 \node (uh2) at (16.0,5.0) [vertexX] {$u_{h,2}$};
 \node (up1) at (23.0,8.0) [vertexX] {$u_{p,1}$};
 \node (w) at (12.0,1.0) [vertexX] {$w$};

{\color{\redC{}}
\draw [->, line width=\redT{}] (uk1) to [out=45, in=110] (up1);
\draw [->, line width=\redT{}] (u11) to [out=70, in=125] (uh1);
\draw [->, line width=\redT{}] (w) to [out=0, in=230] (up1);
\draw [->, line width=\redT{}] (u11) to [out=310, in=180] (w);

\draw [->, line width=\redT{}] (uh1) -- (uk1);
\draw [->, line width=\redT{}] (up1) -- (uh2);
\draw [->, line width=\redT{}] (uh2) -- (uk2);
\draw [->, line width=\redT{}] (uk2) -- (u11);
}

{\color{\blueC{}}
 \draw [->, line width=\blueT{}] (up1) -- (uh1);
 \draw [->, line width=\blueT{}] (uh1) -- (uk2);
 \draw [->, line width=\blueT{}] (uk2) -- (w);
 \draw [->, line width=\blueT{}] (w) -- (uh2);
 \draw [->, line width=\blueT{}] (uh2) -- (uk1);
 \draw [->, line width=\blueT{}] (uk1) -- (u11);

 \draw [line width=\blueT{}] (u11) to [out=90, in=180] (5,15);
 \draw [line width=\blueT{}] (5,15) to [out=0, in=180] (19,15);
 \draw [->, line width=\blueT{}] (19,15) to [out=0, in=90] (up1);

 }
\node at (12,0.5) {\mbox{ }};
\end{tikzpicture} 

  \caption{The \good{} decomposition used in Case 4.3.}\label{figx8}
\end{center}
\end{figure}
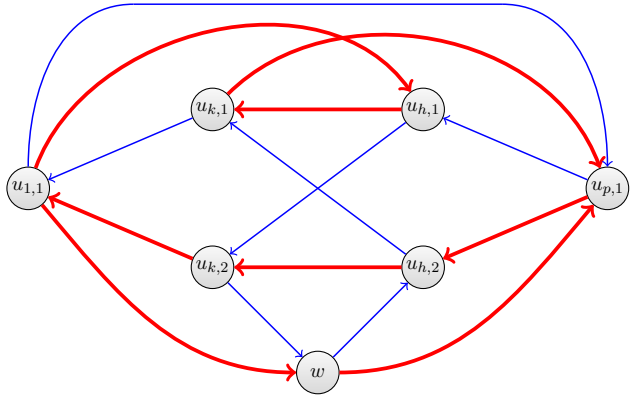

This completes the proof of the theorem.
\end{pf}

 \section{Concluding remarks}\label{sec:disc}

All proofs in this paper are constructive and can be turned into polynomial algorithms for finding \good{} decompositions. Thus, the problem of finding a \good{} decomposition in a semicomplete composition, which has one, admits a polynomial time algorithm. 

Recall that strong semicomplete compositions generalize both strong semicomplete digraphs and strong  quasi-transitive digraphs. However, they do not generalize locally semicomplete digraphs and their generalizations in- and out-locally semicomplete digraphs. A digraph $D$ is {\bf in-locally semicomplete} ({\bf out-locally semicomplete}, respectively) if the in-neighbourhood (out-neighbourhood, respectively) of every vertex of $D$ indices a semicomplete digraph. (For information on in- and out-locally semicomplete digraphs, see e.g.  \cite{bang2009} and \cite[Chapter 6]{bang2018}.)

While there is a characterization of locally semicomplete digraphs having a \good{} decomposition (see Theorem \ref{thm04}), no such a  characterization is known for in-locally semicomplete digraphs\footnote{\GG{Clearly, such a characterization, if it exists, could be easily transformed into that of locally out-semicomplete digraphs.}} and it would be interesting to obtain such a characterization or at least establish the complexity of deciding whether an in-locally semicomplete digraph has a \good{} decomposition. Similar questions are of interest for other generalizations of semicomplete digraphs such as generalizations of quasi-transitive digraphs overviewed in \cite{galeana2018}.

\end{document}